\documentclass[%
 aip,
 amsmath,amssymb,
 reprint,%
]{revtex4-1}

\usepackage{graphicx}
\usepackage{epstopdf}
\usepackage{dcolumn}
\usepackage{bm}

\usepackage[utf8]{inputenc}
\usepackage[T1]{fontenc}
\usepackage{mathptmx}
\usepackage{etoolbox}
\usepackage{epstopdf}
\usepackage{array}
\usepackage{booktabs}
\usepackage{xcolor}
\usepackage{CJKutf8}

\makeatletter
\def\@email#1#2{%
 \endgroup
 \patchcmd{\titleblock@produce}
  {\frontmatter@RRAPformat}
  {\frontmatter@RRAPformat{\produce@RRAP{*#1\href{mailto:#2}{#2}}}\frontmatter@RRAPformat}
  {}{}
}%
\makeatother

\begin{document}

\begin{CJK*}{UTF8}{gbsn}

\preprint{AIP/123-QED}

\title{Trajectory-optimized cluster-based network model for the sphere wake}

\author{Chang Hou (侯昶)}
\affiliation{School of Mechanical Engineering and Automation, Harbin Institute of Technology, Shenzhen 518055, Peoples' Republic of China}

\author{Nan Deng (邓楠)}%
\affiliation{School of Mechanical Engineering and Automation, Harbin Institute of Technology, Shenzhen 518055, Peoples' Republic of China}
\email{dengnan@hit.edu.cn}

\author{Bernd R. Noack}%
\affiliation{School of Mechanical Engineering and Automation, Harbin Institute of Technology, Shenzhen 518055, Peoples' Republic of China}
\affiliation{Hermann-F\"{o}ttinger-Institut,  Technische Universit\"{a}t Berlin, M\"{u}ller-Breslau-Stra{\ss}e 8, D-10623 Berlin, Germany}
\email{bernd.noack@hit.edu.cn}

\date{\today}

\begin{abstract}
We propose a novel trajectory-optimized Cluster-based Network Model (tCNM) for nonlinear model order reduction from time-resolved data following Li \emph{et al.} ["Cluster-based network model,"  J. Fluid Mech. \textbf{906}, A21 (2021)] and improving the accuracy for a given number of centroids.
The starting point is $k$-means++ clustering which minimizes the representation error of the snapshots by their closest centroids. The dynamics is presented by `flights' between the centroids.
The proposed trajectory-optimized clustering aims to reduce the kinematic representation error further by shifting the centroids closer to the snapshot trajectory and refining state propagation with trajectory support points. Thus, curved trajectories are better resolved.
The resulting tCNM is demonstrated for the sphere wake for three flow regimes, including the periodic, quasi-periodic, and chaotic dynamics.
The representation error of tCNM is 5 times smaller 
as compared to the approximation by the closest centroid.
Thus, the error is at the same level as Proper Orthogonal Decomposition (POD) of same order.
Yet, tCNM has distinct advantages over POD modeling: 
it is human interpretable by representing dynamics by a handful of coherent structures and their transitions;
it shows robust dynamics by design, i.e., stable long-time behavior;
and its development is fully automatable, i.e., it does not require tuneable auxiliary closure and other models.
\end{abstract}

\maketitle

\end{CJK*}

\section{Introduction}
\label{sec1}

Reduced-order modeling is a cornerstone of theoretical fluid mechanics and a critical enabler for understanding, 
estimation, dynamic modeling, control, and optimization.
For over one century, a myriad of vortex models 
explain fundamental aspects of coherent structure dynamics
\citep{Pullin1998arfm, Cottet2000book}, e.g., 
the vortex pair behind a cylinder \citep{Foeppl1913} 
and the formation of a vortex street \citep{Karman1911}.
Over hundred years ago, the Galerkin method \citep{Galerkin1915vi} served as a foundation
for computational fluid mechanics \citep{Fletcher1984book} and modern reduced-order models \citep{Holmes2012book}.
In the 50s and 60s, 
\citet{Stuart1971arfm} employed stability modes in the Galerkin framework
for an elegant description of self-amplified amplitude-limited dynamics. 
Another wave of physical Galerkin models starts with this stability theory framework.
Mathematical Galerkin models \citep{Busse1991asr, Noack1994jfm} aim at more general dynamics by employing complete Hilbert space bases.
These modes are typically only available for simple geometries.

The Proper Orthogonal Decomposition (POD), 
particularly the snapshot method \citep{Sirovich1987qam1}, 
opened an increasingly popular avenue to 
data-driven reduced-order modeling \citep{2017_AIAA_Taira_modal},
fueled by the increasing availability of high-quality numerical and experimental data.
\citet{1988_JFM_aubry_dynamics} pioneered POD modeling
for the unforced turbulent boundary layer, 
one of the most complex flows to start with.
An avalanche of POD models followed for a myriad of configurations.
\citet{1998_JFM_podvin_low} proposed a low-dimensional model for the minimal channel flow unit for the purpose of physical understanding. 
She continued to develop an accurate high-dimensional POD model for the wall region of a turbulent channel flow \citep{2009_PoF_podvin_proper}. 
\citet{2009_JCP_Bergmann_enablers} further proposed a pressure extended POD model with no additional cost and tested that on the two-dimensional square cylinder wake.

The first principles based Galerkin model has been completed with a pressure-term representation
 \citep{2005_JFM_noack_need} and modal energy-flow analysis.
The unresolved fluctuations may be incorporated 
with closures as recently reviewed by \citet{San2021pf}.
Non-intrusive Galerkin models avoid any first-principle information.
In this category, Sparse Identification of Nonlinear Dynamics (SINDy) 
aim at human interpretable models \citep{2016_PNAS_brunton_discovering}. 
In addition to the spectrum of dynamical system identification,
also different Galerkin expansions have been developed.
Examples are the balanced POD by \citet{2005_IJBC_rowley_model},
the multiscale-POD by \citet{2019_JFM_Mendez_multi},
the Dynamic Mode Decomposition (DMD) by \citet{Rowley2009jfm,2010_JFM_schmid_dynamic,2016_SIAM_Kutz_dynamic},
and the recursive DMD by \citet{2016_JFM_noack_recursive}.
All Galerkin models tend to be fragile, 
inaccurate for transient dynamics, and 
restricted to a narrow range of operating conditions.
A key reason is the elliptical Galerkin ansatz with global modes neglecting the convective dynamics\citep{2016_JFM_noack_snapshots}.
The advances in machine learning algorithms also attracted widespread concern for data-driven modeling, e.g., with Artificial Neural Networks (ANN) \citep{2019_CNSN_San_artificial,2019_PoF_Zhu_machine}, 
Deep Neural Networks (DNN) \citep{2017_JFM_Kutz_deep}.

A fundamentally different data-centric avenue, 
cluster-based modeling, rapidly gains traction.
While the Galerkin method is based on a linear superposition of modes, the cluster-based approach conceptualizes the dynamics
as `flights' in a state-space which is discretized by select centroids.
This avenue was pioneered by  \citet{2006_CMAME_burkardt_pod}. \citet{2014_JFM_kaiser_cluster} proposed the Cluster-based Reduced-Order Modeling (CROM) method. CROM partitions the flow data into clusters 
and describes the flow dynamics with a Cluster-based Markov Model (CMM). 
The temporal evolution is modeled as 
a probabilistic Markov model of the transition dynamics. 
The state vector of cluster probabilities may initially start in a single cluster but eventually diffuses to a fixed point representing the post-transient attractor. 
CMM  has provided valuable physical insights for the Ahmed body wake \citep{2014_JFM_kaiser_cluster}, combustion-related mixing \citep{2014_EF_cao_cluster}, and the supersonic mixing layer \citep{2020_PoF_li_cluster}.

A challenge for CMM is the temporal evolution:
the state quickly diffuses over the whole attractor, 
often within one typical time period. 
This loss of dynamic information is alleviated by
Cluster-based Network Model (CNM) from time-resolved data
\citep{2021_JFM_li_cluster,2021_SCIA_fernex_cluster}.
In CNM, the dynamics is modeled on a directed network\citep{2013_RSP_Barabasi_network,2008_Elsevier_Arenas_synchronization} 
based on motions between centroids. 
This motion may be pictured in an airport analogy.
The centroids constitute the nodes, like airports, 
and the flights represent the edges, like a deterministic-stochastic flight schedule.
The schedule allows only a few possible flights 
with corresponding probabilities and flight times
consistent with the data.

The CNM can be seen as an extension 
of the CMM 
using the network model 
instead of the standard Markov model to describe the transient dynamics. 
\citet{2022_JFM_deng_cluster} later applied a scale-dependent hierarchical clustering to the CNM under a mean-field consideration and generalized the CNM in the case of multiple invariant sets.
\citet{2019_JFM_nair_cluster} applied clustering as the basis 
for coarse-graining the control law before optimization.

The accuracy of the CNM depends on the centroids.
Clustering minimizes the representation error
of the snapshots to the closest centroids,
but ignores the possibility of linear interpolations (flights).
Evidently, the averaging of the snapshots in a cluster 
leads to centroids where distinct vortex structures may be smoothed out.
In addition, the flights between centroids may not accurately track the real trajectories.
In this study, centroids are shifted closer to the snapshot trajectories, and constraints between centroids are allowed to get an accurate state propagation.
The resulting trajectory-optimized Cluster-based Network Model (tCNM) shows better consistency with the original dataset as compared to the CNM, and leads to a better understanding of the physical mechanisms involved in the flow dynamics.

Table ~\ref{Table_comparison} compares the discussed data-driven reduced-order models.
The starting point is the POD model with a deterministic description in the corresponding subspace, followed by the CROM (CMM) with a stroboscopic temporal prediction of discrete states. The CNM with time-continuous uniform motion on a network of routes between the nodes, and the tCNM with the trajectory optimization based on centroid shift and inter-cluster constraint.
\begin{table}
\caption{Comparison between POD, CROM, CNM and tCNM. For details see text.}
\label{Table_comparison}
\begin{ruledtabular}
\begin{tabular}{ccccc}
Model & POD & CROM & CNM & tCNM \\
\hline
Data driven & \checkmark & \checkmark & \checkmark & \checkmark \\
Time resolved data  &   & \checkmark & \checkmark & \checkmark \\
Intrusive formulation exists & \checkmark &   &   &   \\
Robust long time behavior &   & \checkmark & \checkmark & \checkmark \\
Accurate  long time behavior &   &   & \checkmark & \checkmark \\
Trajectory tracking &   &   &   & \checkmark \\
\end{tabular}
\end{ruledtabular}
\end{table}
In analogy to CFD models, the POD model can be conceptualized as the Large Eddy Simulations (LES) describing the coherent structures with the statistically rare events being underrepresented, and the CROM is similar to the unsteady Reynolds-averaged Navier-Stokes equations describing the transient mean flow. The CNM is seen as the LES with the subgrid-scale vortices being simply modeled, while the tCNM can be regarded as the LES with a more accurate subgrid-scale model.

As demonstration, 
we consider the well-investigated 
three-dimensional incompressible flow around a sphere. 
Despite the simple geometry, 
the wake is fully three-dimensional and capable of featuring complicated vortical structures. 
Many studies have investigated the flow structure in the wake of spheres using dye or smoke visualizations \citep{1937_ENHK_moller_experimentelle,1974_JFM_achenbach_vortex,1990_JFE_sakamoto_study,1999_PoF_leweke_vortex,2013_JFS_chrust_loss} 
and numerical simulations \citep{1999_JFM_johnson_flow,2001_JCP_kim_immersed,2002_JCP_ploumhans_vortex,2006_PoF_yun_vortical,2009_JFM_giacobello_wake,2018_JFM_rajamuni_transverse}.
As the Reynolds number $Re$ increases, 
the sphere wake undergoes a series of bifurcations and transitions on the route to turbulence. 
This transition scenario is an exquisite testing ground for reduced-order modeling. 
In this study, we focus on three representative flow regimes of the sphere wake, including periodic, quasi-periodic, and chaotic flow. 

The manuscript is organized as follows.
Section~\ref{sec2} describes the numerical plant and the computational validation of the simulation.
Section~\ref{sec3} proposes the trajectory-optimized Cluster-based Network Model. 
In section~\ref{sec4}, the detailed flow characteristics of sphere wake at different $Re$ are introduced.
In section~\ref{sec5}, tCNM is performed on the sphere wake featuring three dynamics, including periodic flow, quasi-periodic flow, and chaotic flow, and compared to the CNM.
Section~\ref{sec6} summarizes the main findings and gives some suggestions for future directions.

\section{Flow configuration and numerical method}
\label{sec2}

\subsection{Flow configuration}
\label{sec2.1}

\begin{figure}
\includegraphics[width=7.5cm]{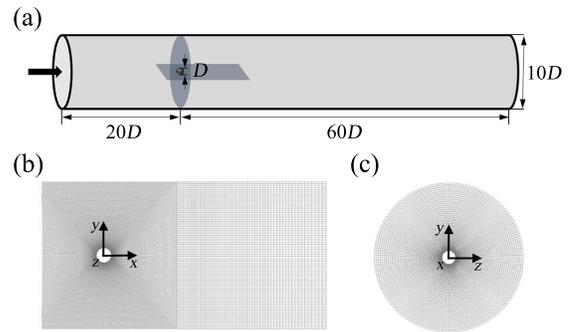}
\caption{(a) Three-dimensional sketch of the flow configuration. $D$ is the diameter of the sphere. (b) Grid expanded view of plane $z=0$. (c) Grid expanded view of plane $x=0$.}
\label{Sketch_of_numerical_simulation}
\end{figure}

The three-dimensional sketch of the flow configuration is presented in figure~\ref{Sketch_of_numerical_simulation} (a). 
A sphere of diameter $D$ is placed in a uniform flow with streamwise velocity $U_{\infty}$.
The computational domain is described in a Cartesian coordinate system where the origin is placed in the center of the sphere, and the $x$-axis is aligned with the streamwise direction.
The whole domain is a cylinder tube, and its sizes in the $x$, $y$, $z$ directions are 80$D$, 10$D$, and 10$D$, respectively. The inlet is 20$D$ upstream from the center of the sphere. 
The choice of these domain parameters ensures a balance of minimal distortion of the flow structures due to the outer boundary conditions and a reasonable computational cost.
The computational domain is discretized into a number of hexahedral grids, figure~\ref{Sketch_of_numerical_simulation} (b) and (c) illustrate the representative expanded view of the grids around the sphere.

Based on the Cartesian coordinates, the net forces acting on the sphere can be decomposed into three components $F_{\alpha }$, where $\alpha  = x, y$ and $z$. These forces are non-dimensionalized as
\begin{equation}
C_\alpha  = \frac{F_\alpha }{1/2\rho U_{\infty}^{2} S},
\end{equation}
where $C_\alpha $ represents the streamwise force coefficient $C_x$, transverse force coefficients $C_y$ of the $y$-axis and $C_z$ of the $z$-axis, 
$\rho$ is the flow density, 
$S = \pi D^2 / 4$ is the projected surface area of sphere in streamwise direction. 
The total drag force coefficient $C_d = C_x$. 
For an axisymmetric sphere, the lift coefficient can be in any direction in the $yz$ plane \citep{2021_JFM_zhao_flow}, and the total lift force coefficient $C_l$ can be defined as
\begin{equation}
C_{l}=\sqrt{C_{y}^{2}+C_{z}^{2}}.
\end{equation}
Furthermore, the Strouhal number $St$ can be written as
\begin{equation}
St=\frac{D f}{U_{\infty}},
\end{equation}
where $f$ is the characteristic frequency.

\subsection{Direct Numerical Simulation}
\label{sec2.2}
The fluid flow is simulated by solving the non-dimensionalized incompressible Navier-Stokes equations,
\begin{eqnarray}
\frac{\partial \bm{u}}{\partial t}+\bm{u} \cdot \nabla \bm{u}+\nabla p-\frac{1}{Re} \nabla^{2} \bm{u}=0,\\
\nabla \cdot \bm{u}=0, 
\end{eqnarray}
where $\bm{u}$ denotes the velocity vector $(u, v, w)$ referred to a cylindrical $(x, r, \theta)$ coordinates system whose origin is placed at the center of the sphere, $p$ is the static pressure, $Re = U_{\infty } D/ \nu$ and $\nu$ is the kinematic viscosity.
The scales for the non-dimensionalization are characteristic length $D$, free stream velocity $U_{\infty }$, time units $D/U_{\infty }$ and pressure scales $\rho U_{\infty}^{2} $.

The commercially available CFD package ANSYS Fluent 15.0 is employed in the present study to discretize the governing equations using the cell-centered Finite Volume Method (FVM). 
The Pressure-Implicit Split-Operator (PISO) algorithm is used to achieve the pressure-velocity coupling. 
The spatial discretization in the governing equations is in the second order. The first-order implicit scheme is used to discretize the temporal term. In addition, small integration temporal steps are selected in order to satisfy the Courant-Friedrichs-Levy (CFL) condition. Thus the Courant number is below 1 for all simulations undertaken.
Note that all the simulations included herein are run at least throughout $t = 400$ non-dimensional time units to ensure enough characteristics of the quasi-periodic or chaotic flow.
Moreover, the data processing is mainly performed using the whole converged range, discarding the first 200 units to avoid any transient stage.

Regarding the boundary conditions, a uniform streamwise velocity is imposed at the inlet boundary, with $\bm{u}= (U_{\infty }, 0, 0)$.
For the outlet, an outflow condition is imposed which implements a Neumann condition for the velocity, 
$\partial_x \bm{u} =(0, 0, 0)$,
and a Dirichlet condition for the pressure, $p = 0$. 
A no-slip boundary condition is imposed at the surface of the sphere. Additionally, slip boundary conditions are set on the cylinder tube walls therefore any wake-wall interpolations are excluded.

\subsection{Validation}
\label{sec2.3}
Grid independence tests are performed at $Re = 300$ to select the appropriate grid size for the numerical study. 
The relative errors of the typical parameters using three grids with different node numbers are compared before conducting extensive simulations.
\newcommand{\tabincell}[2]{\begin{tabular}{@{}#1@{}}#2\end{tabular}}
\begin{table}
\caption{Numerical results at $Re = 300$ using three grids with different node numbers (million). 
$\overline{C_{d}}$, $\overline{C_{l}}$ are the mean values of drag and lift coefficients, respectively. 
$C_{d}^{'}$, $C_{l}^{'}$ are the standard deviations.
$St$ is the Strouhal number. 
The corresponding error deviations in brackets are calculated with respect to the values of the dense grid.}
\label{Table1}
\begingroup
\renewcommand{\arraystretch}{1.1} 
\begin{ruledtabular}
\begin{tabular}{cccccc}
Grids & $\overline{C_d}$ & $C_{d}^{'}$ & $\overline{C_l}$ & $C_{l}^{'}$ & $St$\\[6pt] 
\hline
\tabincell{c}{Coarse\\3.6 } & \tabincell{c}{0.6653\\(0.42$\%$)} & \tabincell{c}{0.001902\\(2.51$\%$)} & \tabincell{c}{0.07045\\(1.37$\%$)} & \tabincell{c}{0.009839\\(3.63$\%$)} & 0.1363\\[5pt] 
\specialrule{0em}{2pt}{2pt}  
\tabincell{c}{Moderate\\5.1 } & \tabincell{c}{0.6631\\(0.10$\%$)} & \tabincell{c}{0.001946\\(0.25$\%$)} & \tabincell{c}{0.06959\\(0.13$\%$)} & \tabincell{c}{0.01012\\(0.89$\%$)} & 0.1363\\[5pt] 
\specialrule{0em}{2pt}{2pt}  
\tabincell{c}{Dense\\8.1 } & 0.6625 & 0.001951 & 0.06950 & 0.01021 & 0.1363 \\[5pt] 
\end{tabular}
\end{ruledtabular}
\endgroup
\end{table}

Table~\ref{Table1} shows the grid independence analysis results with three sets of grids, the coarse grid, the moderate grid, and the dense grid.
For all three sets, the grids are refined near the surface of the sphere. The spacing is increased with a ratio $1.1$, and the first layer is small enough to ensure $y^{+}$ on the sphere surface is less than $1$.
The difference between these three sets mainly comes from the difference in node numbers on the sphere surface and the wake region. 
The mean and standard deviation of the typical parameters of the sphere wake, e.g., lift and drag coefficients and Strouhal number, are compared. 
For this set of variables, there is less than a $1\%$ variation between the moderate and the dense grid, which means further refinement of the grid will only affect the results weakly.
Thus, the moderate grid confirm the adequacy of the numerical results and are adopted in the present study.

The results are compared against the available data from other literature to validate the numerical method. 
\begin{table}
\caption{Validation for the numerical method at $Re = 300$, with compared with the listed literature.}
\label{Table2}
\begin{ruledtabular}
\begin{tabular}{cccc}
Study & $\overline{C_d}$ & $\overline{C_l}$ & $St$\\[5pt] 
\hline
Present Study & 0.663 & 0.069 & 0.136\\
\citet{1999_JFM_johnson_flow} & 0.656 & 0.069 & 0.137\\
\citet{2001_JCP_kim_immersed} & 0.657 & 0.067 & 0.137\\
\citet{2009_JFM_giacobello_wake} & 0.658 & 0.067 &  0.134\\
\citet{2018_JFM_rajamuni_transverse} & 0.665 & 0.070 & 0.137\\
\citet{2021_PoF_Chizfahm_data} & 0.662 & 0.070 & 0.137\\
\end{tabular}
\end{ruledtabular}
\end{table}
Table~\ref{Table2} illustrates the mean drag coefficient, mean lift coefficient, and the Strouhal number, calculated in this study and other literature for $Re = 300$.
The results from different studies are similar to each other. Since the value of these variables is small and sensitive, it can be concluded that the present study has a satisfactory match with them. In addition, the result from the present study is in agreement with the most recent numerical results by \citet{2021_PoF_Chizfahm_data}.

Overall, the comparative validations presented here provide confidence that our computational grid and selected schemes are adequate to resolve the flow characteristics of the sphere wake appropriately, and sufficient for testing the reduced-order modeling method.

\section{Trajectory-optimized Cluster-based Network Model}
\label{sec3}
\begin{figure*}
\includegraphics[width=\linewidth]{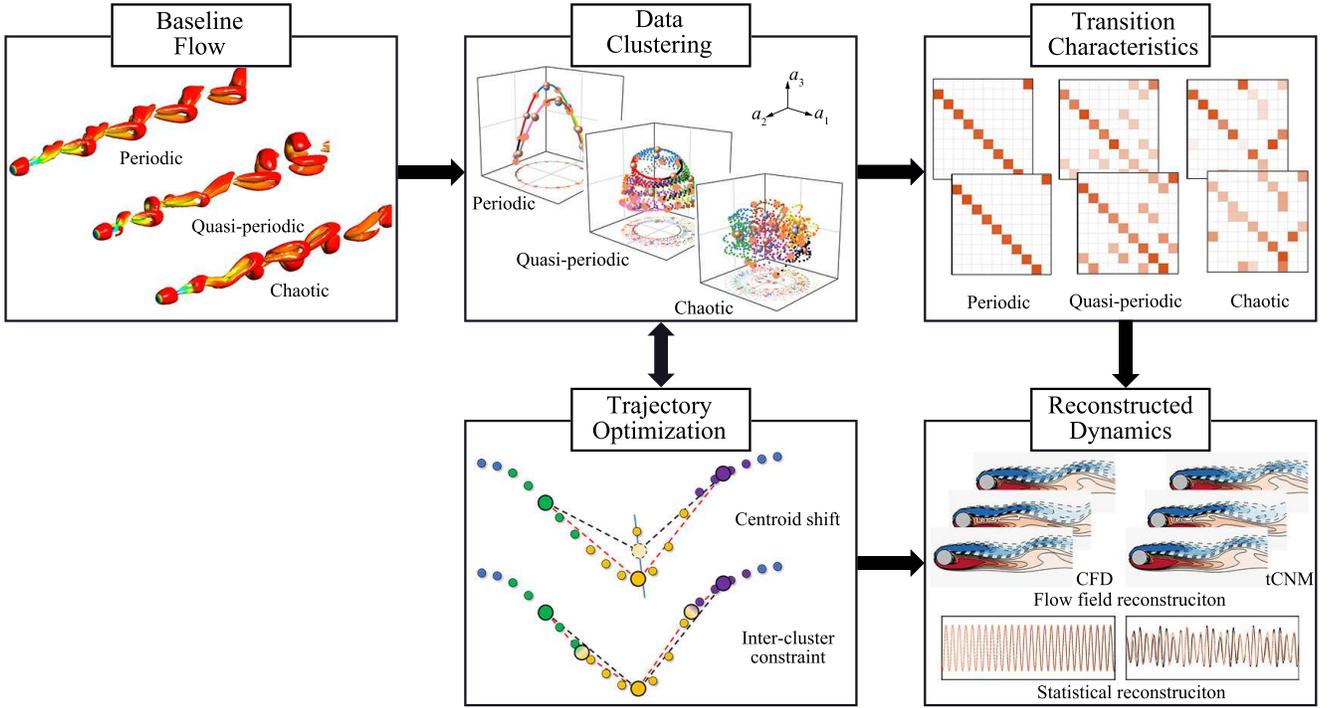}
\caption{Overview of the presented trajectory-optimized Cluster-based Network Model framework, exemplified for three flow regimes of the sphere wake involving periodic, quasi-periodic, and chaotic dynamics.
After clustering, the datasets are grouped into 10 clusters. 
Trajectory optimization procedure is used to shift the centroids and build the trajectory support points,
which are defined as nodes of the network.
The dynamics is presented by `flights' between the nodes based on two matrixes, including transition probability $\mathbf{Q}$ and transition time $\mathbf{T}$ inferred from data.}
\label{Method_framework}
\end{figure*}
In this section, we propose a novel data-driven reduced-order modeling methodology, namely the trajectory-optimized Cluster-based Network Model (tCNM). 
The tCNM is based on the Cluster-based Network Model (CNM) \citep{2021_SCIA_fernex_cluster,2021_JFM_li_cluster} which constitutes a deterministic-stochastic grey-box model for the coherent structure dynamics of complex flow. 
The general framework of tCNM exemplified for the sphere wake is illustrated in figure~\ref{Method_framework}.
The first step is the data collection of baseline flow, where a set of snapshots are collected from experiments or numerical simulations. 
The second step is the identification of the cluster distribution using an unsupervised clustering algorithm to achieve the coarse-graining. 
Following is the automated optimization of the cluster centroids and the building of trajectory support points, which correspond to the nodes of the network. 
This step is a critical improvement of tCNM compared with CNM, with the representation error being reduced. 
The last step is to characterize the motion along with the nodes and defines a smoother, more accurate state propagation trajectory than CNM. 
We discuss the details of the methodology in the following subsections.
The cluster analysis is described in \S~\ref{sec3.1}.
The characterization of the transition dynamics is illustrated in \S~\ref{sec3.2}.
The trajectory optimization strategy used in the network model is discussed in \S~\ref{sec3.3}.
In the end, \S~\ref{sec3.4} defines the validation of the model.

\subsection{Coarse-graining with clustering}
\label{sec3.1}
In this study, clustering is achieved with the unsupervised $k$-means++ algorithm, which groups the data into clusters such that the intra-cluster similarity is maximized and the inter-cluster similarity is minimized \citep{2014_JFM_kaiser_cluster}.
For an ensemble of $M$ statistically representative, time-resolved snapshots $\bm{u}^{m}$, where $m = 1, \ldots , M$, the $k$-means++ algorithm partitions the snapshots into $K$ clusters, defined as $\mathcal{C}_k$, where $k = 1, \ldots , K$.
The snapshots are equidistantly sampled, resolving the coherent structure evolution on  a time window which is statistically representative.
In particular, the time step determines the temporal resolution of the snapshots and the accuracy of the transition dynamics by the cluster-based model.
The resolved time range affects the accuracy of the first and second moments.
In our experience, at least 10 periods of the dominant frequency are required for reasonably accurate statistical moments, 
and at least 10 snapshots per characteristic period 
are needed  to resolve the evolution of the dominant coherent structures.
In this study, snapshots are sampled at a  time step of $\Delta t = 0.2$ over a time window more than 20 periods, with approximately 40 snapshots per period.
The clustering algorithm can also be applied to the low-dimensional representations or observables of the full state snapshots.

The algorithm iteratively updates the optimal location of the cluster centroids according to the data distribution in the state space.
Firstly, the $K$ centroids, namely $\bm{c}_{k}^{0}$ in this study, are chosen from the snapshots. 
Compared to the $k$-means algorithm, the $k$-means++ algorithm optimizes the initial centroids to avoid local optimum and speed up convergence.
Next, each snapshot is affiliated to the closest centroid, following the cluster affiliation function $f$:
\begin{equation}
f\left(\bm{u}^{m}\right)=\arg \underset{k}{\min} \left\|\bm{u}^{m}-\bm{c}_{k}^{0}\right\|_{\Omega},
\end{equation}
where $\left\| \ \right\|_{\Omega}$ denotes the standard norm on a Hilbert space of the domain $\Omega$, which is defined as:
\begin{equation}
\|\bm{u}\|_{\Omega}:=\sqrt{(\bm{u}, \bm{u}) _\Omega} ,
\end{equation}
where the inner product of two square-integrable velocity fields $\bm{u}(x)$ and $\bm{v}(x)$ in domain $\Omega$ reads:
\begin{equation}
(\bm{u}, \bm{v}) _\Omega = \int_{\Omega} \bm{u}(x) \cdot \bm{v}(x) \ \mathrm{d} x .
\end{equation}
After all the snapshots are allocated, the intra-cluster variance $J$ of this cluster partition is computed as:
\begin{equation}
J\left(\bm{c}_{1}^{0}, \ldots, \bm{c}_{K}^{0}\right)=\sum_{k=1}^{K} \sum_{\bm{u} ^{m} \in \mathcal{C}_{k}}\left\|\bm{u} ^{m}-\bm{c}_{k}^{0}\right\|_{\Omega}^2.
\end{equation}
Next, the original cluster centroids $\bm{c}_{k}^{0}$ are updated by averaging all the snapshots within the cluster $\mathcal{C}_k$:
\begin{equation}
\bm{c}_{k}^{0}=\frac{1}{n_{k}} \sum \bm{u}^{m}, \quad \bm{u}^{m} \in \mathcal{C}_{k}, 
\end{equation}
where $n_{k}$ is the number of snapshots in the cluster $\mathcal{C}_k$. 
The snapshots are allocated to clusters again, according to the new $K$ centroids. 
These processes will be repeated until the intra-cluster variance $J$ is minimized below a specified tolerance. 
As the initial cluster centroids are generated randomly, $k$-means++ algorithm is repeated 100 times in this study to ensure an optimal partition.
The averaged centroids $\bm{c}_{k}^{0}$ and the cluster affiliation $k(\bm{u}^{m})$ of snapshots within the corresponding clusters $\mathcal{C}_k$ are taken as the final results.

The computational cost of clustering algorithms is also considered for large high-dimensional datasets.
An optional pre-processing can be done with a lossless POD method to conduct data compression \citep{2014_JFM_kaiser_cluster,2021_JFM_li_cluster,2021_SCIA_fernex_cluster,2022_JFM_deng_cluster}.
The snapshot $\boldsymbol{u}^{m}$ can be exactly expressed by the POD expansions as:
\begin{equation}
\bm{u}^{m}(\bm{x}) - \bm{u}_{0}(\bm{x}) \approx \sum_{i=1}^{N} a_{i}^{m}\bm{u}_{i}(\bm{x}),
\end{equation}
where $\bm{u}_{0}$ is the constant mean flow field, 
$\bm{u}_{i}(\bm{x})$ are the spatial POD modes, with the corresponding mode coefficients $\bm{a}^{m} = \left[a_{1}^{m}, \ldots, a_{N}^{m}\right]$ for temporal evolution. 
A complete basis for the decomposition can be expected when $N = M$\citep{1993_ARFM_Berkooz_proper}.
Thus, the distance can be computed with $\bm{a}^{m}$ instead of the full-dimensional snapshots, which reads as:
\begin{equation}
d\left(\bm{u}^{m}, \bm{u}^{n}\right) = \left\|\bm{u}^{m}-\bm{u}^{n}\right\|_{\Omega}=\left\|\bm{a}^{m}-\bm{a}^{n}\right\|.
\end{equation}

The computational load of clustering is significantly reduced with dimensionality reduction, e.g., an optional POD pre-processing.
The clustering algorithm is applied to the $N$-dimensional state vector $\bm{a}^{m}$ instead of performing an expensive computation on the full domain with millions of grid nodes.
This computational saving of POD pre-processing can be further amplified according to the number of iterations required for a $k$-means algorithm until convergence \citep{2021_SCIA_fernex_cluster}.  
We emphasize that the POD is only an optional pre-processing to make clustering computationally feasible 
and is not necessary for the clustering algorithm itself.
In this work, we compress the time-resolved snapshots with a \emph{lossless} POD pre-processing, where the leading $500$ POD modes can resolve more than $99.9\%$ of the fluctuation energy.
Thus, the  model performance is not affected by POD pre-processing.
For the small chosen time step, 
CNM and tCNM resolve only the non-trivial transitions between the centroids.
Thus, the rapid  diffusion of states in the original cluster-based Markov models\citep{2014_JFM_kaiser_cluster} is avoided.

\subsection{Characterizing the transition dynamics}
\label{sec3.2}
Since tCNM and CNM share similar transition dynamics, before we detail the trajectory optimization procedure, CNM is briefly reviewed.

In CNM, the datasets are grouped into clusters by the clustering algorithm,
and cluster centroids are considered as network nodes.
Moreover, the motion between these nodes is characterized by a network model with transition probabilities and transition times.

The nonlinear dynamics is reconstructed as a linear transition between centroids.
The transitions between clusters rely on the direct transition matrix, which is typically modeled with a first-order Markov model. The probability of moving from cluster $\mathcal{C}_j$ to $\mathcal{C}_k$ is described by the direct transition matrix $ \mathbf{Q}=Q_{k j} \in R^{K \times K}$, which is deducted from the snapshots as:
\begin{equation}
Q_{k j} = \frac{n_{k j}}{n_{j}},
\end{equation}
where $n_{k j}$ is the number of snapshots moving from cluster $\mathcal{C}_j$ to $\mathcal{C}_k$,
$n_j$ is the number of transitions departing from $\mathcal{C}_j$ regardless of the destination cluster.
Moreover, the direct transition ignores inner-cluster residence probability with $Q_{k j}=0$ when $k=j$.
The clusters are renumbered so that the first cluster has the highest state probability,
the second cluster is chosen with the highest transition probability from the first one,
and so on until all are sorted \citep{2014_JFM_kaiser_cluster}. 
Note that the matrix is uniformed such that all elements of this matrix are non-negative, and the elements of each column sum up to unity.

\begin{figure}
\includegraphics[width=7.5cm]{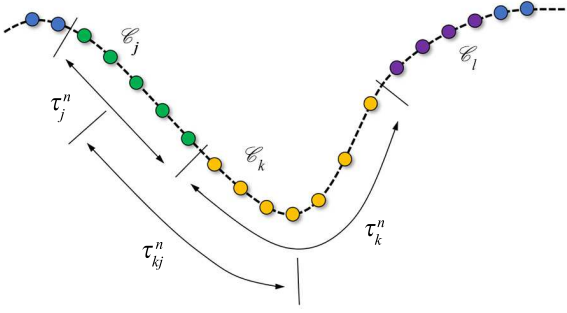}
\caption{Transition from cluster $\mathcal{C}_j$ to $\mathcal{C}_k$. Different colors correspond to different clusters. $\tau _{j}^{n}$ and $\tau _{k}^{n}$ are the residence time in cluster $\mathcal{C}_j$ and $\mathcal{C}_k$, respectively. $\tau _{kj}^{n}$ is the individual transition time.}
\label{Transition_time}
\end{figure}

The transition times are the other transition property, let $t^n$ be the time of the first snapshots to enter and $t^{n+1}$ be the last snapshots to leave during one individual transition in cluster $\mathcal{C}_j$.
The residence time $\tau _{j}^{n} = t^{n+1} - t^{n}$ corresponds to the duration of staying in cluster $\mathcal{C}_j$ at this transition. 
For the transition time, we further define the individual transition time $\tau _{kj}^{n}$ from cluster $\mathcal{C}_j$ to $\mathcal{C}_k$ as half the residence time of both clusters, as shown in figure~\ref{Transition_time}.
Averaging $\tau _{kj}^{n}$ from all individual transitions, the transition time is expressed as:
\begin{equation}
T_{k j}=\frac{\sum_{n=1}^{n_{k j}} \tau_{k j}^{n}}{n_{k j}},
\end{equation}
where $n_{k j}$ is the number of individual transitions from $\mathcal{C}_j$ to $\mathcal{C}_k$ from all the transitions. 
Similar to the direct transition matrix $\mathbf{Q}$, we can also build the transition time matrix $ \mathbf{T} = T_{k j} \in R^{K \times K}$. 

Based on the transition probabilities and the transition times, we assume a uniform state propagation between two centroids $\bm{c}_{j}^{0}$ and $\bm{c}_{k}^{0}$ as:
\begin{equation}
\bm{u}^{m}(\bm{x},t) = \alpha_{k j}(t)\, \bm{c}_{k}^{0} (\bm{x})+\left( 1-\alpha_{k j}(t) \right)\, \bm{c}_{j}^{0} (\bm{x}),
\label{Eqn:FlightCNM}
\end{equation}
where $\alpha_{k j}(t) = \left(t-t_{j}\right) / T_{k j}$ , $t_j$ is the time when the centroids $\bm{c}_{j}^{0}$ is left.

\subsection{Trajectory optimization}
\label{sec3.3}
The proposed improvement relates to the uniform straight-line flight \eqref{Eqn:FlightCNM} of CNM.
First, the trajectory from centroid $\bm{c}_j^0$ to $\bm{c}_k^0$ will be curved, i.e., enter cluster $\mathcal{C}_k$ on a different point than predicted by \eqref{Eqn:FlightCNM}.
Second, a typical trajectory may not even be near $\bm{c}_k^0$.
In the tCNM, we replace the straight-line flights from cluster $\mathcal{C}_j$ to $\mathcal{C}_k$ by a curved trajectory through a shifted centroid $\bm{c}_j$, an averaged entry point $\bm{r}_{kj}$, and a shifted centroid $\bm{c}_k$.
The shifted centroid $\bm{c}_k$ is on the line connecting $\bm{c}_k^0$ and the average of all entry points and all exit points to and from cluster $\mathcal{C}_k$.
The shift amplitude is chosen to reduce the representation error of all trajectories through cluster $\mathcal{C}_k$.
In other words, the shifted centroid $\bm{c}_k$ is employed for all trajectories regardless of the departure cluster $\mathcal{C}_j$ while the support point $\bm{r}_{kj}$  is chosen for a specific cluster transition from $\mathcal{C}_j$ to $\mathcal{C}_k$.
We emphasize a shift in the clustering paradigm:  
The original centroids optimize the average squared distance of the snapshots to their nearest centroids.
In contrast, the shifted centroids reduce the distance from the snapshots to the corresponding approximation of the trajectory.
In the following, the construction is detailed.

\begin{figure}
\includegraphics[width=7.5cm]{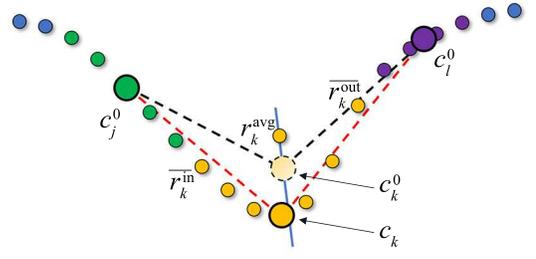}
\caption{Centroid shift. 
The cluster affiliation is displayed as in figure \ref{Transition_time}.
For cluster $\mathcal{C}_k$, a straight line is generated based on $\bm{r}_{k}^{\mathrm{avg}}$ and $\bm{c}_{k}^{0}$,
where $\bm{r}_{k}^{\mathrm{avg}}$ is the average of all entry and exit snapshots.
The averaged centroid $\bm{c}_{k}^{0}$ is shifted to $\bm{c}_k$,
which is defined by the closest snapshot from the straight line.
The black dashed line is the propagation with $\bm{c}_{k}^{0}$ of CNM, the red dashed line is the propagation with the shifted centroid $\bm{c}_k$ of tCNM.
}
\label{Centroid_shift}
\end{figure}

For simplicity, 
we first explain the centroid shift for a single trajectory, as illustrated in figure~\ref{Centroid_shift}.
A curved trajectory can be found for a transition from cluster $\mathcal{C}_j$ to $\mathcal{C}_k$ and then to $\mathcal{C}_l$. 
CNM uniformly propagates the state through the motion between the three original centroids $\bm{c}_{j}^{0}$, $\bm{c}_{k}^{0}$ and $\bm{c}_{l}^{0}$, as shown by the black dashed line. 
The state propagation shows significant deviation from the curved trajectory, because the centroid $\bm{c}_{k}^{0}$ from the mean operator erases lots of critical information and is not suitable to represent the states in $\mathcal{C}_k$.
The centroid is shifted to be as close to the trajectory as possible, ensuring that the error between the state propagation and the snapshots is minimized.
This optimization directly benefits kinematic state propagation by reducing trajectory variance rather than intra-cluster variance.
We use $\bm{c}_k$ to denote the shifted centroids to distinguish them from the original centroids $\bm{c}_k^0$.

The centroid shift is inferred from the the snapshots.
The entry and exit points in cluster $\mathcal{C}_k$ are defined as:
\begin{equation}
\begin{aligned}
    \overline{\bm{r}_{k}^{\mathrm{in}}}= \frac{1}{n_{c}} \sum \bm{u}^{m}, \quad \bm{u}^{m} \in \mathcal{C}_{k}  \ \& \  \bm{u}^{m-1} \notin \mathcal{C}_{k} ,\\
    \overline{\bm{r}_{k}^{\mathrm{out}}}= \frac{1}{n_{c}} \sum \bm{u}^{m}, \quad \bm{u}^{m} \in \mathcal{C}_{k}  \ \& \  \bm{u}^{m+1} \notin \mathcal{C}_{k} ,
\end{aligned}
\label{Average_trajectories}  
\end{equation}
where $n_{c}$ is the number of trajectory segments in $\mathcal{C}_{k}$.
Hence, $\overline{\bm{r}_{k}^{\mathrm{in}}}$ and $\overline{\bm{r}_{k}^{\mathrm{out}}}$ are the averages of inlet and outlet snapshots of $\mathcal{C}_k$, respectively. 
The midpoint of two boundary averages of cluster $\mathcal{C}_k$ reads:
\begin{equation}
\bm{r}_{k}^{\mathrm{avg}}=\frac{\overline{\bm{r}_{k}^{\mathrm{in}}}+\overline{\bm{r}_{k}^{\mathrm{out}}}}{2}. 
\end{equation}
A straight line $\bm{l}_{k}$ passing through $\bm{r}_{k}^{\mathrm{avg}}$ and the original centroid $\bm{c}_{k}^{0}$ in Euclidean geometry is given by the two point form of line equation in Euclidean geometry:
\begin{equation}
\bm{l}_{k}(\alpha) = \alpha\, \bm{r}_{k}^{\mathrm{avg}} + \left( 1-\alpha \right)\, \bm{c}_{k}^{0}, \quad \alpha \in \mathbb{R}. 
\end{equation}
The most representative state for a curved trajectory can be identified at the junction with the line $\bm{l}_{k}$.
For cluster with multiple trajectories, the shifted centroid is defined as the average of all their junctions.
With the above geometric method, the shifted centroid is close to the extreme in a curved trajectory and the average in a straight trajectory.
The trajectory optimization by shifting centroids to more representative states is data-driven and generalizable, achieving the minimum deviation for state motion on trajectories.

For a trajectory consisting of discretized snapshots, the junction is the nearest snapshot to the straight line, with the distance read as:
\begin{equation}
D_{\bm{l}_k}^{m} = \min_{\bm{u}^{m}\in \mathcal{C}_{k}} \left\|\bm{u}^{m} - \bm{l}_{k}(\alpha^\prime) \right\|_{\Omega}, 
\end{equation}
where $ \alpha^\prime = (\bm{u}^{m} - \bm{c}_{k}^{0}, \bm{r}_{k}^{\text {avg}} - \bm{c}_{k}^{0}) _\Omega / \left\| \bm{r}_{k}^{\text {avg}} - \bm{c}_{k}^{0} \right\|_{\Omega}^2 $ consider the foot of the perpendicular on the line $\bm{l}_{k}$ from $\bm{u}^{m}$.
The shifted centroid can hence be obtained by averaging these chosen snapshots from all the trajectories.
%

The second trajectory optimization is introducing inter-cluster constraint for the transitions between two clusters.
Since a cluster can have multiple destination clusters, a refined state motion according to the cluster transitions will provide more accurate state propagation.
\begin{figure}[h]
\includegraphics[width=7.5cm]{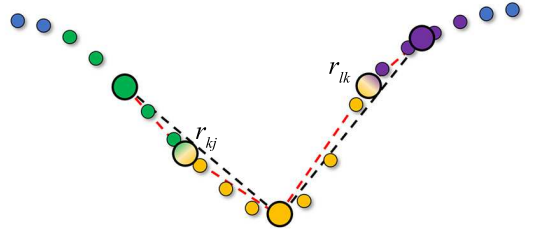}
\caption{Inter-cluster constraint. The cluster affiliation is displayed as in figure \ref{Transition_time}. $\bm{r}_{kj}$, $\bm{r}_{lk}$ are the trajectory support points of the adjacent clusters. 
The red dashed line is the direct propagation between cluster centroids, black dashed line is the optimized propagation considering the trajectory support points.}
\label{Inter-cluster_constraint}  
\end{figure}
Figure~\ref{Inter-cluster_constraint} illustrates the sketch of inter-cluster constraint for a single trajectory.
Trajectory support points are introduced to ensure the proximity of an accurate enter to the destination cluster.

The average entry point
 from cluster $\mathcal{C}_j$ to $\mathcal{C}_k$
 is noted as  
trajectory support points $\bm{r}_{kj}$ 
 and defined by
\begin{equation}
    \bm{r}_{kj} 
    = \left \langle \bm{u}^{m} \right\rangle_{j\mapsto k}.
\label{trajectory_support}  
\end{equation}
Here, $\left \langle \bm{u}^{m} \right\rangle_{j\mapsto k}$
averages over all
snapshots on segment
crossing the boundary between cluster $j$ and cluster $k$.
These snapshots satisfy
$\bm{u}^m \in \mathcal{C}_{k}
\quad  \bm{u}^{m-1} \in \mathcal{C}_{j}$
or 
$\bm{u}^{m} \in \mathcal{C}_{j}  
\quad  \bm{u}^{m+1} \in \mathcal{C}_{k}
$.
The model trajectory connects 
$\bm{c}_j$, the
entry point $\bm{r}_{kj}$
and $\bm{c}_k$
and can be expected to be more accurate than the original straight-line segment.

In CNM, the transition between cluster $\mathcal{C}_j$ and $\mathcal{C}_k$ is based on a uniforn motion between two centroids.
In tCNM, the model trajectory
is a spline through centroids
and `border-crossing' trajectory support points.
The times from centroid $\bm{c}_j$ to $\bm{r}_{kj}$  
from $\bm{r}_{kj}$
 to $\bm{c}_k$
 are estimated to be $T_{kj}^-$
 and $T_{kj}^+$ respectively.
 In other words, 
 the transition time
 is decomposed in two parts,
 $T_{kj} = T_{kj}^- + T_{kj}^+$.
 
An efficient realization of CNM requires POD of the snapshot data.
The clustering adds a small computational cost 
in comparison to building POD, 
about 10\% for this study.
The computational load of the transition model, i.e., building
the matrixes $\mathbf{Q}$ and $\mathbf{T}$, and integrating the model is small even compared to the cost of clustering.
Details on the computational load of these algorithms are provided in the previous study by \citet{2021_SCIA_fernex_cluster}.
In this study, the computational costs of CNM and tCNM
add 11\% and 30\%, respectively, to the cost of POD, taken as 100\%.
tCNM is more costly than CNM because of the matrix operations to identify $\bm{c}_k$ as well as to get $\bm{r}_{kj}$.

\subsection{Validation}
\label{sec3.4}

There are usually two kinds of prediction errors for cluster based models: from the spatial perspective, the representation error due to the insufficient representation by cluster centroids, and from the temporal perspective, the transition error due to the inaccurate reconstruction of complex transition relationships. 
Considering CNM, for complex phase-space trajectories, more accuracy will be achieved by the high-order Markov chains \citep{2021_SCIA_fernex_cluster}.
The high-order CNM requires more previous states with time delays to decide the destination of the current state, enabling finer differentiation of transition dynamics and lower transition error. 
However, our focus is on decreasing representation error with a trajectory-optimized extension, thus, the traditional first-order model is chosen in this study.
The trajectory optimization by centroid shift and inter-cluster constraint can be seen as a procedure to reduce the representation error without producing extra transition error.
It is worth noting that there will be no coupling relationship between these two kinds of errors once the number of clusters is determined. 

The chosen number of clusters $K$ significantly affects the prediction error.
The network model with a small number of clusters can capture the dominant behavior of transition dynamics. Fewer transitions between clusters will be accounted for, enabling easier interpretation of the results with physical insights \citep{2021_SCIA_fernex_cluster}. Therefore the transition error is relatively small. 
However, a larger representation error is induced due to representing the dynamics by fewer clusters.
The details of snapshots are eliminated as the centroids average the snapshots in each cluster.
In contrast, a large $K$ will allow CNM to model the transition dynamics with more detail, which means a more accurate representation of snapshots.
However, considering more clusters, the transition relationship will be more complex.
Especially with an excessively large $K$, the centroids are too dense so that the nearly identical motions are clustered separately into adjacent trajectories, and even random switching between trajectories is introduced. In this case, the resulting CNM exhibits non-physical transitions, leading to a significant increase in transition error.
Thus, we expect a sweet spot with optimal prediction error based on good representation error and an accurate prediction for the number of clusters. 
\citet{2019_JFM_nair_cluster} determined an optimal choice of cluster number by the F-test considering the ratio of inter-cluster variance to the total variance.
A similar balancing strategy, like the elbow method \citep{2014_JFM_kaiser_cluster}, can also be used to get an appropriate cluster number.
In this study, the snapshots are coarse-grained into $K = 10$ clusters, which is consistent with many previous studies\citep{2014_JFM_kaiser_cluster,2019_JFM_nair_cluster,2021_JFM_li_cluster,2020_PoF_li_cluster,2020_PoF_tan_cavity}. This value is in good agreement with the F-test and elbow method, and is large enough to resolve the main transition mechanism.

To statistically evaluate the performance of the reduced-order models, including the representation error and the transition error.
Several metrics are presented: the autocorrelation function, the cluster probability vector and the function of representation error. 

The autocorrelation function avoids the problem of comparing two trajectories directly with finite prediction horizons due to phase mismatch \citep{2021_SCIA_fernex_cluster}. 
Especially for chaotic flow, where a slight change in initial conditions leads to completely different trajectories, thus, the direct comparison of time series is often pointless. Note that for pure periodic flow, the direct comparison of time series often makes sense since there is no phase mismatch. 
The function is defined as\citep{2015_JFM_Protas_optimal}:
\begin{equation}
    R(\tau)=\frac{1}{T-\tau} \int_{0}^{T-\tau} (\bm{u}(\bm{x},t) \cdot \bm{u}(\bm{x},t+\tau))_{\Omega} \ \mathrm{d} t, \tau \in[0, T]. 
\end{equation}

The cluster probability vector is defined as:
\begin{equation}
    \bm{p}=\left[p_{1}, \ldots, p_{k}\right],
\end{equation}
where $p_k$ represents the probability of being in cluster $\mathcal{C}_k$, which is the cumulative residence time normalized by the simulation time horizon, as:
\begin{equation}
    p_{k}=\frac{\sum \tau_{k}^{n}}{T}.
\end{equation}
The vector $\bm{p}$ indicates whether the predicted trajectories could populate the phase space similar to the original data. 
The comparison, to some extent, reveals the transition error of the modeling method.

The representation error of the reconstructed dynamics is defined as:
\begin{equation}
E_r = \frac{1}{M} \sum_{m=1}^M D_{\mathcal{T}}^{m},
\end{equation}
where $D_{\mathcal{T}}^{m}$ is defined as the minimal distance from the snapshot $\bm{u}^m$ to the states on the reconstructed trajectory $\mathcal{T}$:
\begin{equation}
D_{\mathcal{T}}^{m} = \min_{\bm{u}^{n}\in \mathcal{T} } \left\|\bm{u}^{m}-\bm{u}^{n}\right\|_{\Omega}. 
\end{equation}
%

\section{Flow features at different Reynolds number}
\label{sec4}

Flow past a sphere is a prototype wake of bluff bodies commonly encountered in many engineering applications, for instance, the design of drones, air taxis, or micro-robots.
Due to the geometric simplicity and symmetry, the sphere wake has been well investigated for its three-dimensional flow characteristics,
which exhibits different patterns at different $Re$, such as steady, periodic, quasi-periodic, and chaotic flow.
These different kinds of baseline flow are ideal for testing the performance of reduced-order modeling. We can get the dataset of general flow dynamics without changing the numerical configuration but simply adjusting $Re$.

As $Re$ increases, the wake behind a sphere undergoes a series of bifurcations and transitions on its way to chaos.
The flow separation around a sphere is widely accepted to occur at $Re \approx 20$ \citep{2013_JFS_chrust_loss}. 
\begin{figure*}
\centering
\includegraphics[width=\linewidth]{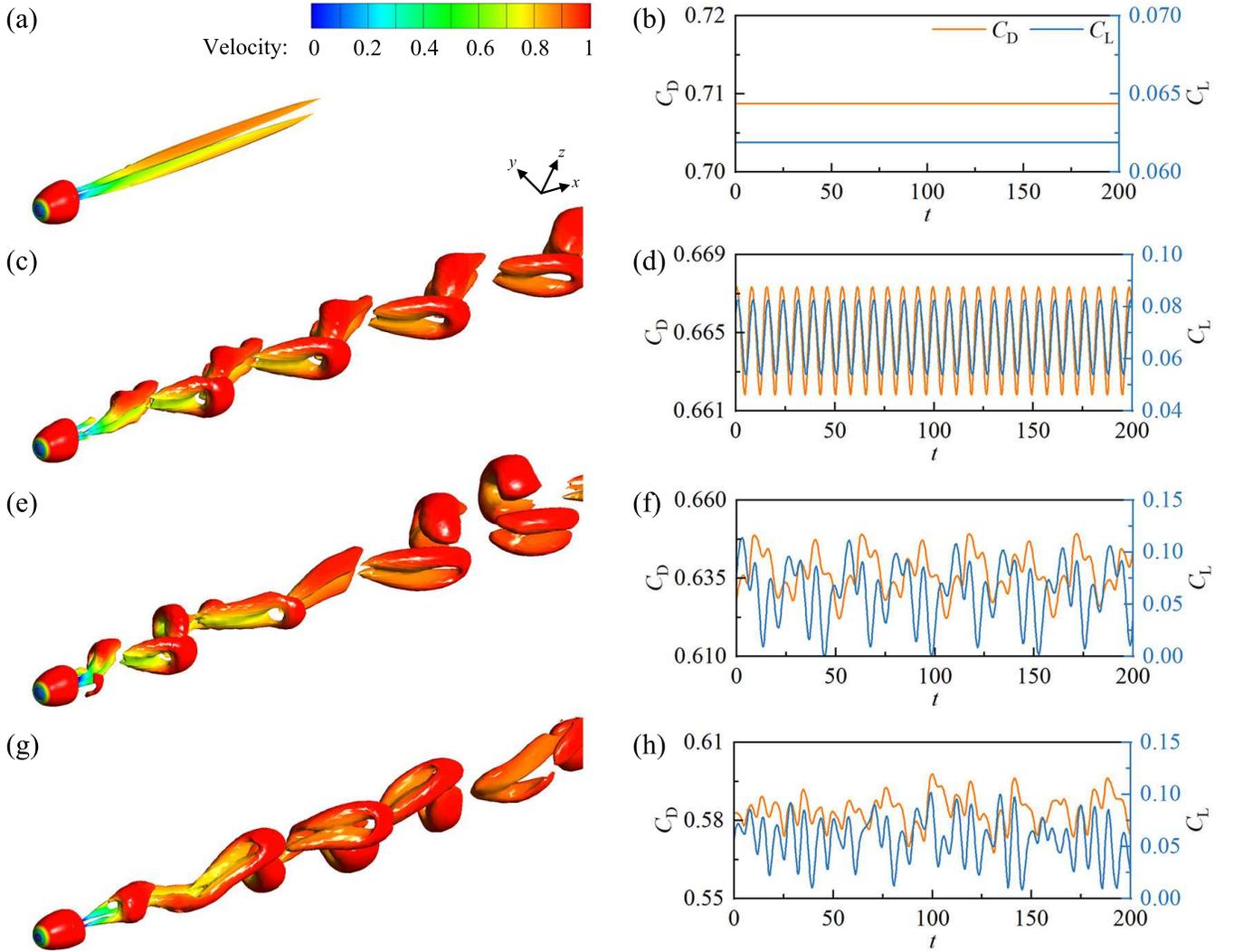}
\caption{Representative vortex structures and temporal evolution of $C_d$, $C_l$ at different $Re$.
The vortices are visualized using iso-surfaces of the $Q$-criterion ($Q$ = 0.001) color-coded with non-dimensional velocity $U/U_{\infty }$, and the visualization of all baseline flows uses the same legend and view.
(a), (b) Planar symmetry steady flow at $Re = 250$, with $C_d$ and $C_l$ being constants.
(c), (d) Periodic flow at $Re = 300$, with periodic shedding of hairpin vortices and a symmetry plane ($xy$ plane).
(e), (f) Quasi-periodic flow at $Re = 350$, with a newly appeared lower frequency modulates the shedding.
(g), (h) Chaotic flow at $Re = 450$, with the irregular shedding and the azimuthal rotation of the separation point, as characterized by chaotic features of $C_d$ and $C_l$.}
\label{Vortex_structure} 
\end{figure*}
The resulting wake remains axisymmetric until $Re \approx 210$, where a pitchfork bifurcation occurs, which breaks the axial symmetry and leads to a steady non-axisymmetric wake consisting of a pair of steady streamwise vortices that are symmetrical about a longitudinal plane \citep{2008_PoF_fabre_bifurcations}, like in figure~\ref{Vortex_structure} (a).
For better observation, a rotation along the $x$-axis is applied to keep the symmetry plane (if present) coincident with the $xy$ plane.
A Hopf bifurcation is observed at $Re \approx 270$, corresponding to the appearance of vortex streets or the shedding of hairpin vortices, like in figure~\ref{Vortex_structure} (c).
The symmetry of the wake with respect to the longitudinal plane still preserves \citep{2008_PoF_fabre_bifurcations}.
At higher $Re \approx 350$, 
the planar symmetry of vortex shedding starts to lose its trace, and the hairpin vortices exhibits differences among each shedding period \citep{2021_JFM_zhao_flow}.
A lower frequency appears and modulates the shedding, e.g., the flow exhibits a longer periodicity of about seven shedding periods at $Re = 350$, which results from a secondary Hopf bifurcation, as shown in figure~\ref{Vortex_structure} (e).
The successive bifurcations eventually lead to a fully chaotic flow at $Re > 420$ \citep{2018_JFM_pan_wake}, revealing more complex vortical structures with three-dimensional properties, as shown in figure~\ref{Vortex_structure} (g).

In this study, the four representative baseline flow regimes mentioned above are investigated, including the planar symmetry steady state at $Re = 250$, the periodic, quasi-periodic, and chaotic flow at $Re = 300$,  $350$, and $450$, respectively. 
The three-dimensional vortex structures and $C_d$, $C_l$ of the sphere are shown in the right column of figure~\ref{Vortex_structure}.
For the planar symmetry steady flow, the streamwise vortices remain stationary, with two threads of opposite axial vorticity, and $C_d$ and $C_l$ are all constants. 
Since the vortex intensity of the corresponding two vortices is not the same, $C_l$ would not be zero but is one order of magnitude smaller than $C_d$.
For periodic flow, the shedding of hairpin vortices dominates the whole flow field, which remains identical among each shedding period with $C_d$ being periodic, while the vortex intensity of the corresponding vortex pair still exhibits a slight difference from each other. Thus, the average $C_l$ would not be zero, but its amplitude will remain periodic.
For quasi-periodic flow, the vortex intensity of the corresponding vortex pair exhibits a larger difference. The similarity of the hairpin vortices among each shedding period is lost, and a newly appeared lower frequency modulates the shedding, with $C_d$ and $C_l$ becoming quasi-periodic.
For chaotic flow, the shedding of hairpin vortices becomes fully irregular, and displays different orientations, with an azimuthal rotation of the separation point at the rear of the sphere\citep{2020_JFM_Lorite_description}. Meanwhile, the wake exhibits obvious lateral oscillations and secondary vortices\citep{2019_JFM_eshbal_measurement}, as characterized by chaotic features of $C_d$ and $C_l$.

In the following section, tCNM will be applied to the nonlinear dynamics of the periodic, quasi-periodic, and chaotic regimes. 

\section{Modeling the sphere wake}
\label{sec5}

In this section, tCNM is applied to periodic (\S~\ref{sec5.1}), quasi-periodic (\S~\ref{sec5.2}) and chaotic (\S~\ref{sec5.3}) flow regimes. 
We evaluate the performance of tCNM by comparing the temporal evolution, autocorrelation functions, and representative vortex distribution profiles against CNM.

At higher $Re$,  coherent structures can still dominate the flow  and be resolved by centroids.
The small-scale stochastic component is filtered out in the clustering process\citep{2021_SCIA_fernex_cluster}.
In contrast,  cluster-based modeling of transient flows from the steady solution 
to post-transient dynamics requires careful preparation of the simulation data and thorough post-processing\citep{2022_JFM_deng_cluster}.

\subsection{Periodic flow}
\label{sec5.1}

\begin{figure}
\includegraphics[width=7.5cm]{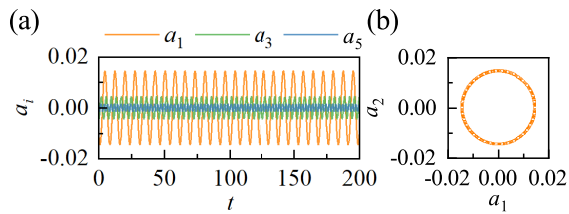}
\caption{Temporal evolution (a) and phase diagram (b) of the leading POD coefficients for periodic flow at $Re = 300$.}
\label{Re300_POD_coefficients}  
\end{figure}

Periodic unsteady flow is characterized by a single frequency and spatial-temporal symmetry.
The flow regime involves only a Hopf bifurcation, and the dynamic reconstruction is simple to implement.
The dataset is firstly compressed with a lossless POD pre-processing. The temporal evolution of the POD mode coefficients is visualized in figure~\ref{Re300_POD_coefficients} (a).
Only the first, third and fifth POD mode coefficients $a_1$, $a_3$ and $a_5$ are presented, as the POD modes usually appear in pairs, and the leading modes can be enough to represent most of the flow characteristics. 
The first pair of POD modes $a_1$, $a_2$ are an order of magnitude larger than all the other coefficients, which is common in periodic flows because the coherent structure dominates and holds most of the kinematic energy.
The phase diagram based on $a_1$, $a_2$ exhibits a standard cycle because of the first harmonic, as viewed in figure~\ref{Re300_POD_coefficients} (b).
Higher-order modes present higher harmonics and even a mixture of different frequencies, which usually do not affect the macro-scale flow characteristics but have an impact on the micro-scale characteristics.

\begin{figure}
\includegraphics[width=7.5cm]{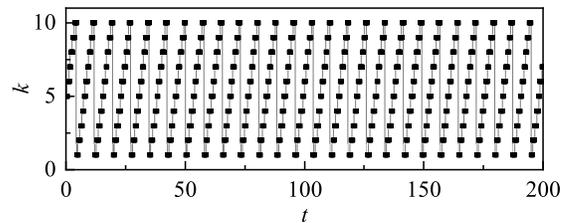}
\caption{Cluster affiliation for periodic flow at $Re = 300$.}
\label{Re300_cluster_affiliation} 
\end{figure}

The clustering is performed on the vectors composed of the leading 500 POD coefficients.
The assignment of each snapshot to the clusters is shown in figure~\ref{Re300_cluster_affiliation}.
The cluster indexes are reordered as \S~\ref{sec3.2} illustrates. 
For periodic flow, the vortex shedding is resolved by the ten clusters, and a clear periodicity can be seen from the cluster affiliation. 
The whole shedding period is divided by ten different vortex patterns corresponding to different clusters, these vortex patterns are further represented by the cluster centroids of CNM and the shifted centroids by tCNM, which are severed as the nodes of the network.
The almost uniform residence time in each cluster shows a strict periodicity and indicates that the size of each cluster is relatively homogeneous.

\begin{figure}
\includegraphics[width=7.5cm]{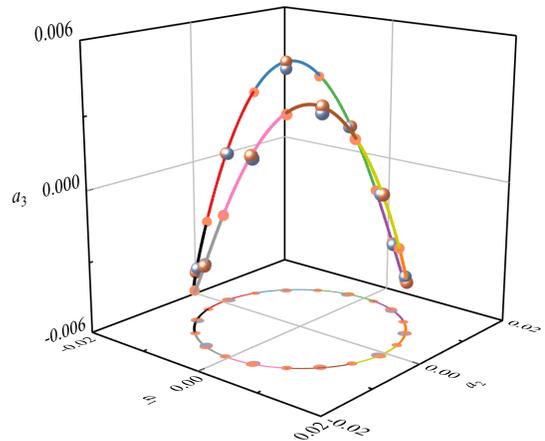}
\caption{Three-dimensional proximity map for periodic flow at $Re = 300$. 
Gray balls are the original CNM centroids, 
orange balls are the shifted tCNM centroids, 
solid circle are the trajectory support points. 
$a _{1}$, $a _{2}$, $a _{3}$ are the corresponding features of the three-dimensional space by MDS, and a projection are conducted into the $a _{1}a _{2}$ plane. The difference between the shifted centroids and the original centroids is comparatively small for the periodic regime with respect to the other two cases in the following study.}
\label{Re300_cluster3d}
\end{figure}

\begin{figure}
\includegraphics[width=7.5cm]{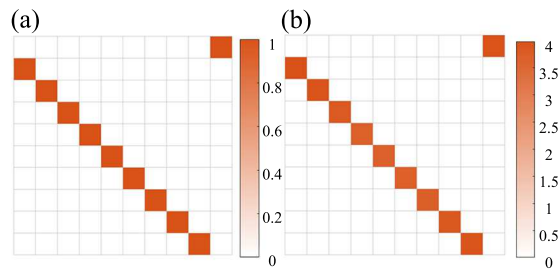}
\caption{Transition matrixes for periodic flow at $Re = 300$. (a) Direct transition matrix $\mathbf{Q}$. (b) Transition time matrix $\mathbf{T}$.}
\label{Re300_transition_probability}
\end{figure}
\begin{figure}
\includegraphics[width=7.5cm]{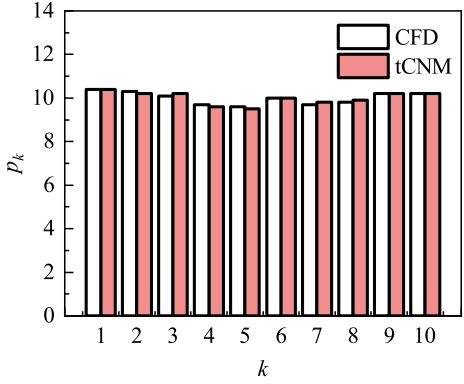}
\caption{Cluster probability distribution for periodic flow at $Re = 300$. Only tCNM is exhibited as CNM and tCNM share the same probability distribution.}
\label{Re300_cluster_probability} 
\end{figure}
\begin{figure}
\includegraphics[width=7.5cm]{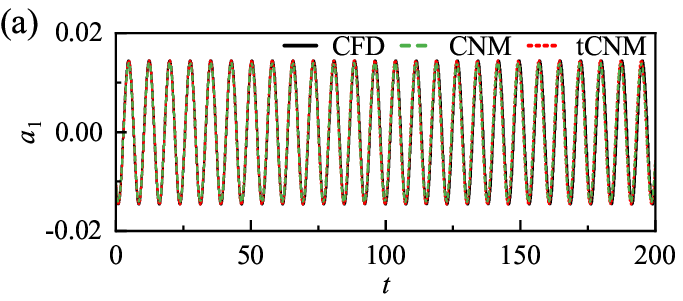}
\includegraphics[width=7.5cm]{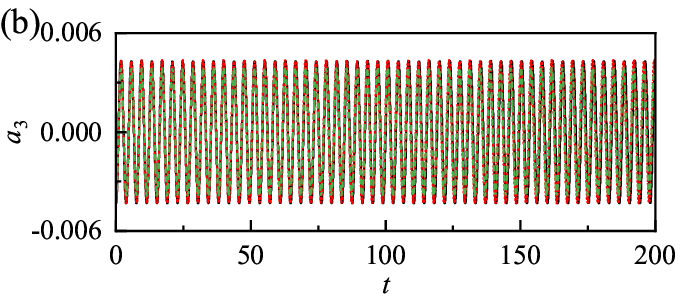}
\includegraphics[width=7.5cm]{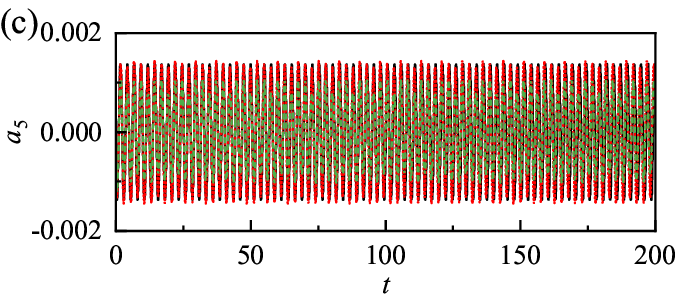}
\caption{Temporal evolution of POD mode coefficients for periodic flow at $Re = 300$. (a) Coefficient $a_1$. (b) Coefficient $a_3$. (c) Coefficient $a_5$. The black solid curve is CFD data, the green dashed curve is the reconstructed temporal dynamics by CNM, the red dotted curve is the reconstructed temporal dynamics by tCNM.}
\label{Re300_coefficients_restore}
\end{figure}
\begin{figure*}
\includegraphics[width=\linewidth]{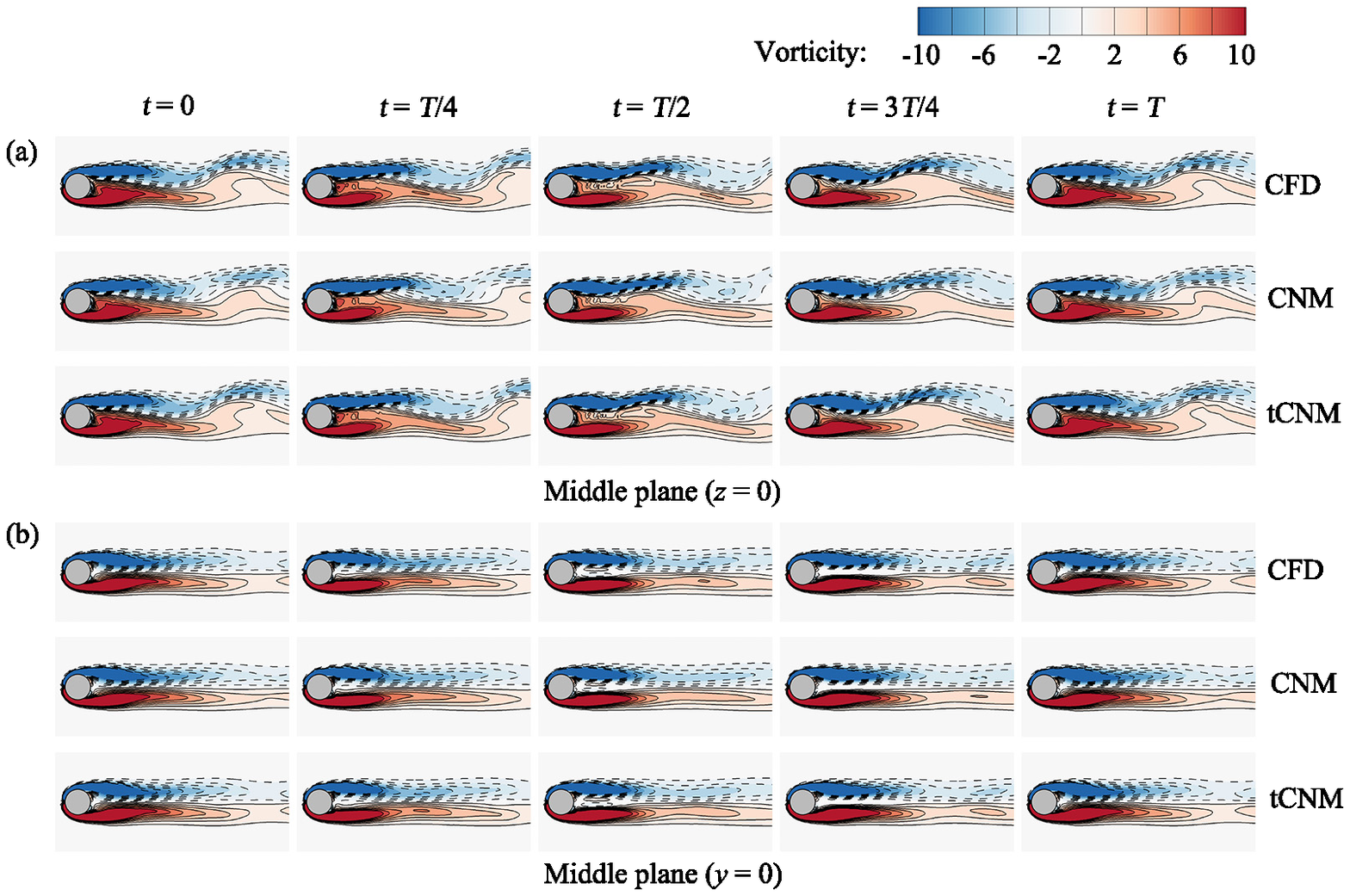}
\caption{Vortex distribution in one typical vortex shedding period of $xy$ plane at $z=0$ (a) and $xz$ plane at $y=0$ (b) for periodic flow at $Re = 300$. 
For the $xy$ plane, the distribution is color-coded with $z$ vorticity, while for the $xz$ plane with $y$ vorticity.
The first row is the CFD data, the second row is the reconstructed dynamics by CNM, and the third row is by tCNM.}
\label{Re300_contours} 
\end{figure*}

The high-dimensional dataset and their cluster affiliation can be easily interpreted by multi-dimensional scaling (MDS), which can project the high-dimensional feature space into a low-dimensional one and then build a proximity map \citep{2020_PoF_li_cluster,2020_PoF_tan_cavity}.
In this study, we project all the POD coefficient vectors of the snapshots and the corresponding nodes, including the original cluster centroids $\bm{c}_{k}^{0}$ of CNM, the shifted cluster centroids $\bm{c}_k$  and the trajectory support points $\bm{r}_{kj}$ of tCNM, into a three-dimensional feature, as shown in figure~\ref{Re300_cluster3d}.
The gray balls represent $\bm{c}_{k}^{0}$, the orange balls represent $\bm{c}_k$, and the solid circle represent $\bm{r}_{kj}$.
The distance between any two dots or balls in the feature space simply represents their similarity level.
These snapshots, represented by dots and colored by their cluster affiliation, form a clear orbit.
A clear difference between CNM and tCNM in the building of nodes can be seen from the proximity map. 
For typical clusters, $\bm{c}_{k}^{0}$ usually exhibits certain deviation from the original trajectory, while $\bm{c}_k$ and the trajectory support points $\bm{r}_{kj}$ are exactly on the trajectory,
making the propagation optimized substantially.
Thus, the propagation of tCNM based on both $\bm{c}_k$ and $\bm{r}_{kj}$ is more accurate with less prediction error, as compared to CNM based on $\bm{c}_{k}^{0}$.

The direct transition matrix $\mathbf{Q}$ and the transition time matrix $\mathbf{T}$ can reveal the temporal evolution of all the possible transitions.
For periodic flow, the flow characteristics are fully resolved, as these matrixes briefly illustrate the limit cycle in one shedding period, as shown in figure~\ref{Re300_transition_probability}. 
The transition matrix shows a strictly periodic transition route, which indicates that the transition dynamics is deterministic and periodic, as also revealed by the cluster affiliation shown in figure~\ref{Re300_cluster_affiliation}.
Each entry of the direct transition matrix $\mathbf{Q}$ is unity, from cluster $\mathcal{C}_k = 1$ to $\mathcal{C}_k = 2$ then gradually to $\mathcal{C}_k = 10$ and finally return to $\mathcal{C}_k = 1$. 
The transition times of all the transitions are similar, revealing that the idling time of each cluster is also similar. Thus, the population probability of these clusters will be uniform.

The temporal dynamics of CNM and tCNM are built based on the transition matrixes $\mathbf{Q}$ and $\mathbf{T}$, as shown in figure~\ref{Re300_cluster_probability}. 
We choose the same initial cluster $\mathcal{C}_k$ as the CFD data and integrate the dynamics over 200 time units.
As the two models share the same transition matrixes, the same temporal evolution are used to ensure a clear comparison for their representation error.
Thus, the cluster population distribution of CNM and tCNM will be the same, we further compare that with the CFD data, as shown in figure~\ref{Re300_cluster_probability}.
For periodic flow, the probability distribution of tCNM (or CNM) is similar to the CFD data, which means the baseline flow is probabilistically well reconstructed with minor transition error. 
The distribution of each cluster is nearly $10\%$, which is consistent with the cluster affiliation in figure~\ref{Re300_cluster_affiliation}. 

Figure~\ref{Re300_coefficients_restore} shows the temporal evolution of POD mode coefficients $a_1$, $a_3$ and $a_5$. The black solid curve represents the original CFD data, the green dashed curve represents CNM and the red dotted curve represents tCNM.
Following \citet{2021_SCIA_fernex_cluster} and \citet{2021_JFM_li_cluster}, the temporal evolution is smoothed by spline interpolation. 
For $a_1$, the two modes all track well the phase, while as for the amplitude, tCNM performs slightly better near the extreme, practically the same as the CFD data.
For the other two modes associated with higher harmonics, the information around the extremes cannot be fully resolved by the averaged centroids $\bm{c}_{k}^{0}$ in CNM, leading to some distortion in the reconstructed dynamics.
However, the distortion can be easily avoided in tCNM using the optimized centroids.

The evolution of vortical structures in one vortex shedding period is illustrated through the slices of 3D flow field, which brings a more intuitive view of the difference between the two models.
Here, the vortex distribution on the middle $xy$ plane and $xz$ plane are visualized in figure~\ref{Re300_contours}.
The top row is the CFD data, the middle row is the reconstructed CNM results, and the bottom row is the tCNM results. 
The results of CNM and tCNM are equidistantly sampled with a time step $T/4$ from $t_0 = 100$, where $T$ is one shedding period detected from the CFD data.
We do these in a similar way for quasi-periodic and chaotic flow.
The corresponding shedding period of the CFD data is chosen for comparison. Since there is no phase mismatch in periodic flow, the CFD data is also chosen from $t_0 = 100$ with the same time step. 
The two models perform well on the first pair of modes for periodic flow, which catch most of the flow characteristics. Therefore, the contour of the two models is all extremely close to the CFD data for most of the snapshots, exhibiting strong applicability of cluster-based reduced-order modeling methods for periodic flow. 
tCNM shows its advantages for the snapshots of $t = T/2$, where the snapshots are at the critical flow state of one shedding period, or in other words, at the moment when one vortex is about to separate from the sphere.
CNM lost certain micro-scale vortex structures due to its inaccuracy in high-order modes, while tCNM can better represent the original data and retain most of the details of the vortex.
As this moment is the critical time point of the shedding period, the importance of accurate reconstruction near this moment is self-evident. Thus, the improved strategy with trajectory optimization in tCNM can be essential.
In addition, for the case where both models can reconstruct the original data accurately, tCNM performs slightly better, resolving more details than CNM due to the impact of inter-cluster constraint, which provides a refined transition reconstruction on the trajectory.

\subsection{Quasi-periodic flow}
\label{sec5.2}

\begin{figure}
\includegraphics[width=7.5cm]{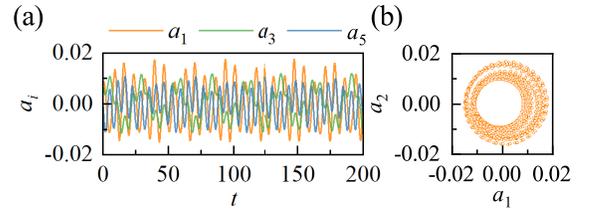}
\caption{
Same as figure~\ref{Re300_POD_coefficients}, but for quasi-periodic flow at $Re = 350$. (a) Phase diagram. (b) The leading POD coefficients.}
\label{Re350_POD_coefficients} 
\end{figure}
For quasi-periodic flow, a newly appeared lower frequency modulates the shedding of hairpin vortices.
Figure~\ref{Re350_POD_coefficients} illustrates the temporal evolution of the POD mode coefficients $a_1$, $a_3$, $a_5$ and the phase diagram of $a_1$, $a_2$.
The mode coefficients exhibit a long periodicity in the time evolution with around seven shedding periods and the phase diagram shows a torus.
Moreover, the similar order of magnitude of the three coefficients indicates that the leading pair of modes cannot completely resolve the hairpin vortices in the quasi-periodic regime.
These differences bring challenges to the accuracy of reduced-order modeling approaches, especially for the representation error of cluster-based methods.

\begin{figure}
\includegraphics[width=7.5cm]{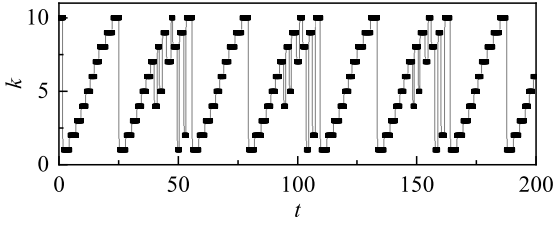}
\caption{
Same as figure~\ref{Re300_cluster_affiliation}, but for quasi-periodic flow at $Re = 350$.}
\label{Re350_cluster_affiliation}
\end{figure}
The assignment of each snapshot to the clusters is shown in figure~\ref{Re350_cluster_affiliation}.
A longer period is observed, which contains two kinds of transition evolution.
In the first part ($t \approx 0$ to $25$), the ten clusters are visited sequentially, and their residence times are almost uniform. In the second part ($t \approx 25$ to $60$), the transition dynamics is more complex, regularly returning to the visited clusters.
This long-period dynamic includes around seven shedding periods, and one shedding period is resolved by a combination of only three or four clusters rather than all the clusters. The different periodic sheddings within a complete loop are resolved by different sorts of clusters.
Consequently, more snapshots are involved in one cluster than periodic flow, which means the representation error will be larger for CNM.

\begin{figure}
\includegraphics[width=7.5cm]{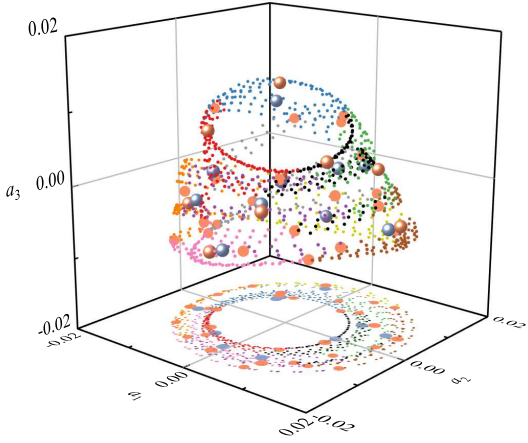}
\caption{
Same as figure~\ref{Re300_cluster3d}, but for quasi-periodic flow at $Re = 350$.}
\label{Re350_cluster3d} 
\end{figure}
Figure~\ref{Re350_cluster3d} shows the three-dimensional view of all the snapshots and propagation nodes of quasi-periodic flow.
The distribution of snapshots resembles the side surface of a cone.
Different parts of the side surface correspond to different clusters depicted in different colors. 
The dynamics appear to be driven by two physical phenomena: a cyclic behavior and a quasi-periodic component that forces the limit cycle to experience cycle-to-cycle variations \citep{2014_EF_cao_cluster}. 
The averaged cluster centroids $\bm{c}_{k}^{0}$ (gray balls) are inside the surface, while the shifted cluster centroids $\bm{c}_k$ (orange balls) and the trajectory support points $\bm{r}_{kj}$ (solid circle) are right on the surface.
This indicates that the averaged centroids are out of the data manifold, while the optimized centroids are closer to the data, thus having a smaller representation error.

\begin{figure}
\includegraphics[width=7.5cm]{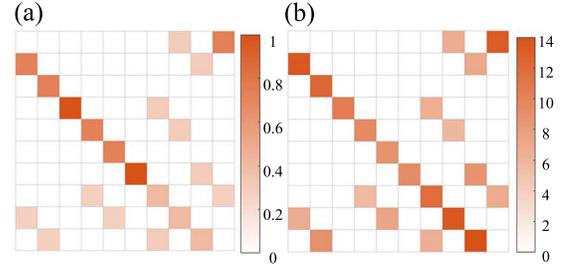}
\caption{
Same as figure~\ref{Re300_transition_probability}, but for quasi-periodic flow at $Re = 350$. (a) Direct transition matrix $\mathbf{Q}$. (b) Transition time matrix $\mathbf{T}$.}
\label{Re350_transition_probability} 
\end{figure}
The direct transition matrix $\mathbf{Q}$ and the transition time matrix $\mathbf{T}$ of quasi-periodic flow are illustrated in figure~\ref{Re350_transition_probability}.
An inspection of $\mathbf{Q}$ reveals that most of the clusters have at least two destination clusters.
However, the cluster transition with higher probabilities form a main loop from cluster $\mathcal{C}_k = 1$ to $\mathcal{C}_k = 10$, which indicates a relatively deterministic transition appeared in figure~\ref{Re350_cluster_affiliation}.
In addition, there are some transitions with minor probabilities, which will lead to random transitions appearing in the network model and introduce the additional transitions errors. The high-order network model is considered as an effective way to solve this problem \citep{2021_SCIA_fernex_cluster}.
From $\mathbf{T}$, the transition time of each cluster in the main loop is different but presents a smooth variation.
That is, the residence time in the adjacent clusters along the trajectory varies continuously, as well as the corresponding vortex structure.

\begin{figure}
\includegraphics[width=7.5cm]{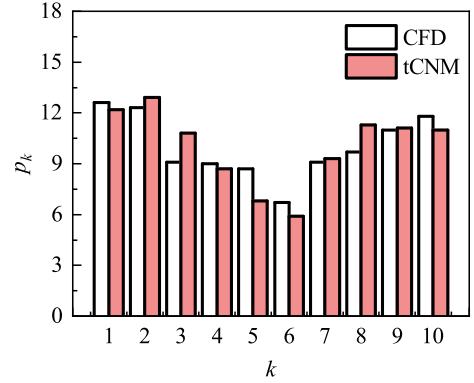}
\caption{
Same as figure~\ref{Re300_cluster_probability}, but for quasi-periodic flow at $Re = 350$.}
\label{Re350_cluster_probability}
\end{figure}
The probability distribution of quasi-periodic flow is well reconstructed by tCNM and CNM, as shown in figure~\ref{Re350_cluster_probability}.
The probability distribution of each cluster is not uniform, which is reasonably larger for the cluster with higher transition probability and longer transition time.
The difference between the distribution of the two models and the CFD data become larger, as compared to the periodic case.
It is reasonable from the following two aspects. One is the transition error becoming larger for quasi-periodic flow. The other is the assumed constant transition time $T_{kj}$ for all trajectories from cluster $\mathcal{C}_j$ to $\mathcal{C}_k$ is a crude assumption due to the coarse-graining of the clustering. The transition times can vary by a large factor and can thus give rise to significant systematic error \citep{2021_JFM_li_cluster}.

\begin{figure}
\includegraphics[width=7.5cm]{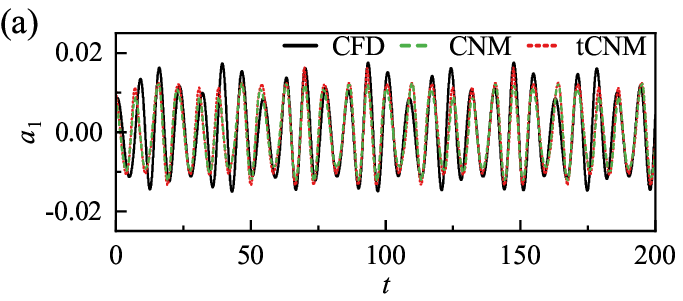}
\includegraphics[width=7.5cm]{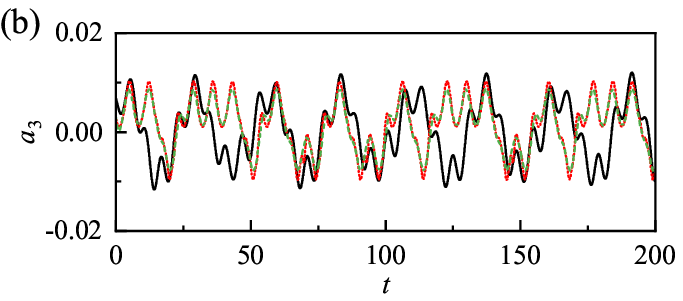}
\includegraphics[width=7.5cm]{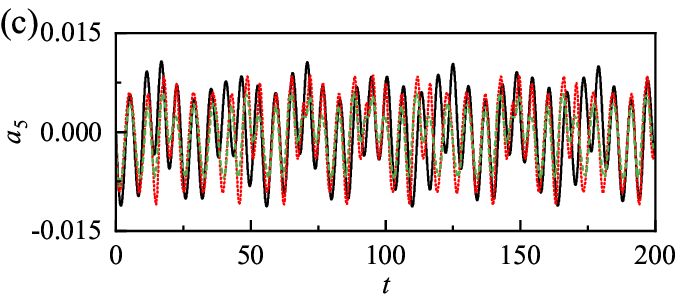}
\caption{
Same as figure~\ref{Re300_coefficients_restore}, but for quasi-periodic flow at $Re = 350$. (a) Coefficient $a_1$. (b) Coefficient $a_3$. (c) Coefficient $a_5$.}
\label{Re350_coefficients_restore}
\end{figure}
\begin{figure}
\includegraphics[width=7.5cm]{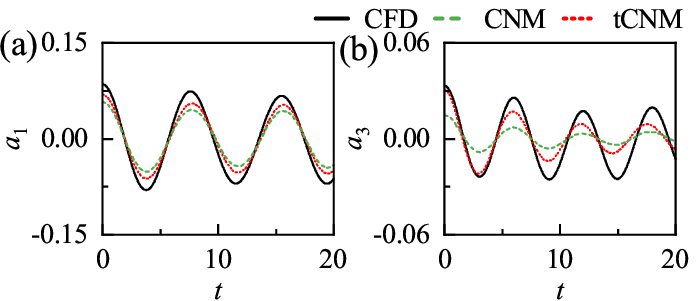}
\includegraphics[width=7.5cm]{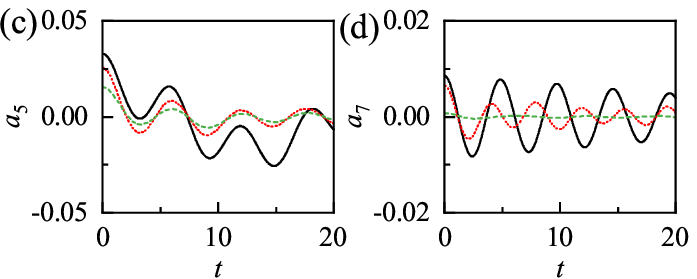}
\caption{Autocorrelation function of POD mode amplitudes for quasi-periodic flow at $Re = 350$. (a) Coefficient $a_1$. (b) Coefficient $a_3$. (c) Coefficient $a_5$. (d) Coefficient $a_7$. The black solid curve is CFD data, the green dashed curve is the CNM data, the red dotted curve is the tCNM data.}
\label{Re350_auto_correlation}
\end{figure}
\begin{figure*}
\includegraphics[width=\linewidth]{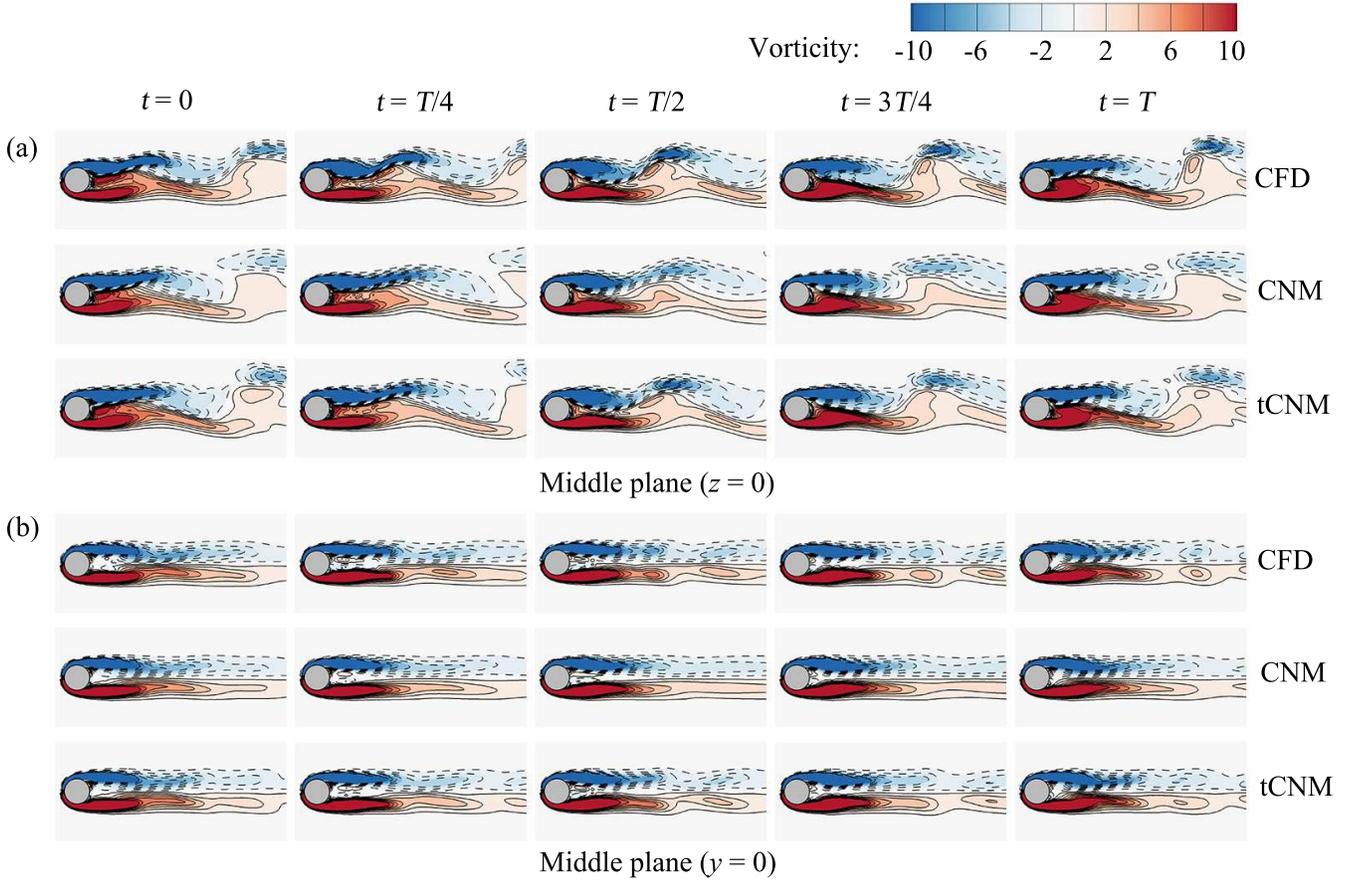}
\caption{
Same as figure~\ref{Re300_contours}, but for quasi-periodic flow at $Re = 350$. (a) $xy$ plane at $z=0$. (b) $xz$ plane at $y=0$.}
\label{Re350_contours}
\end{figure*}
The temporal evolution of POD mode coefficients $a_1$, $a_3$ and $a_5$ is illustrated in figure~\ref{Re350_coefficients_restore}.
For CNM, as the vortex shedding period is highly coarse grained, the loss of the critical flow states by the averaged centroids correspondingly gets more visible than the periodic flow. Therefore the amplitude of the restored coefficients is much smaller.
Despite the phase mismatch due to the complex transition relationship of several clusters, tCNM is shown to be more accurate for restoring the critical dynamics of the quasi-periodic flow 

Considering the impact of phase mismatch, the autocorrelation function of POD mode coefficients $a_1$, $a_3$, $a_5$ and $a_7$ is calculated and shown by figure~\ref{Re350_auto_correlation}.
Similar to the temporal evolution, the black solid curve represents the original CFD data, the green dashed curve represents CNM and the red dotted curve represents tCNM.
Note that a small range of $t$ by definition also makes sense for the comparison. To show the difference between the two modes more clearly, we limit the range of $t$ to 20 time units. For larger $t$, situations are similar. 
The fluctuation level of CNM is much lower than tCNM, due to the information loss. Moreover, the higher-order mode is, the lower the fluctuation level becomes, which is consistent with the periodic flow.
The spatial-temporal resolution of cluster-based models depends on the number of clusters. 
A handful of clusters can well resolve the dominant frequency, but a significantly larger number of clusters is needed to resolve the higher harmonics with lower amplitudes.
This is apparent in figure~\ref{Re350_auto_correlation}, where the dominant dynamics of leading modes is well captured by only ten centroids, but the high-frequency dynamics of higher-order modes is compromised.

The vortex distribution on the middle $xy$ and $xz$ plane are illustrated in figure~\ref{Re350_contours}.
As expected, the most visible difference still occurs at the snapshots of $t = T/2$, as the contour of CNM is more like an averaged flow state rather than an instantaneous one. Several coherent structures disappear, and slight deformation can be seen on the vortex pair. While for tCNM, many of the details are still well retained.
As the propagation of CNM is generally based on only three cluster centroids for one vortex shedding period for quasi-periodic flow, extensive information loss of one centroid will distort the whole trajectory. Thus, for the other snapshots, the inaccuracy of CNM can also be seen. 
Note that the first and last snapshots of tCNM also exhibit certain differences with CFD data. We infer it is mainly due to the phase mismatch of the nearby shedding periods.

\subsection{Chaotic flow}
\label{sec5.3}

\begin{figure}
\includegraphics[width=7.5cm]{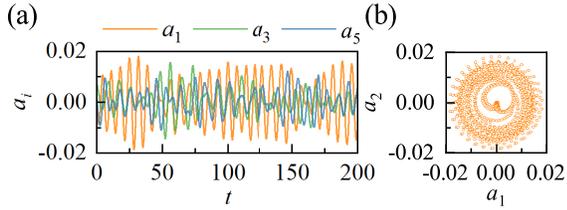}
\caption{
Same as figure~\ref{Re300_POD_coefficients}, but for chaotic flow at $Re = 450$. (a) Phase diagram. (b) The leading POD coefficients.}
\label{Re450_POD_coefficients}
\end{figure}

For chaotic flow, the vortex shedding becomes irregular and fully three-dimensional chaotic, which is explained as the consequence of the azimuthal rotation of the separation point and lateral oscillations of the vortex shedding.
Figure~\ref{Re450_POD_coefficients} shows the temporal evolution of the POD mode coefficients $a_1$, $a_3$, $a_5$ and the phase diagram of $a_1$, $a_2$ for the fully chaotic three-dimensional flow at $Re=450$.
The three coefficients are in the same order of magnitude. 
However, the temporal evolution becomes fully irregular for each coefficient, and no obvious periodic pattern can be found. 
The phase diagram also manifests chaotic characteristics. 
As the temporal evolution of these coefficients gets more complex in chaotic cases, the inter-cluster constraint of tCNM will exhibit more obvious effects in the dynamic representation.

\begin{figure}
\includegraphics[width=7.5cm]{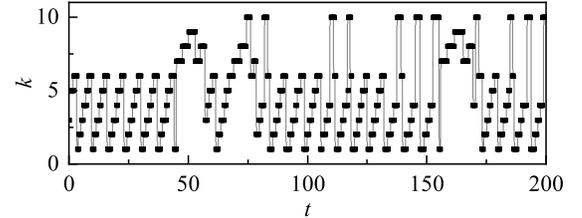}
\caption{
Same as figure~\ref{Re300_cluster_affiliation}, but for chaotic flow at $Re = 450$.}
\label{Re450_cluster_affiliation}
\end{figure}
Figure~\ref{Re450_cluster_affiliation} illustrates the assignment of each snapshot to the clusters.
It is hard to find stable cycling from the cluster affiliation for chaotic flow.
One vortex shedding period is typically resolved by the first six clusters, which is unstable and quickly jumps into the rest four clusters randomly.
The chaotic shedding still retains some characteristics of periodic flow but the incoherent fluctuations and stochastic dynamics dominates.
However, when clustering the chaotic flow, slightly different vortex structures derived from lateral oscillations and secondary vortexes are grouped into the same clusters, resulting in an increase in the intra-cluster variance and a decrease in the representation ability of the cluster centroids.
This introduces a more significant extra representation error.
This kind of representation error is relatively small in periodic or quasi-periodic flow, as the coherent structures dominate in these flows.

\begin{figure}
\includegraphics[width=7.5cm]{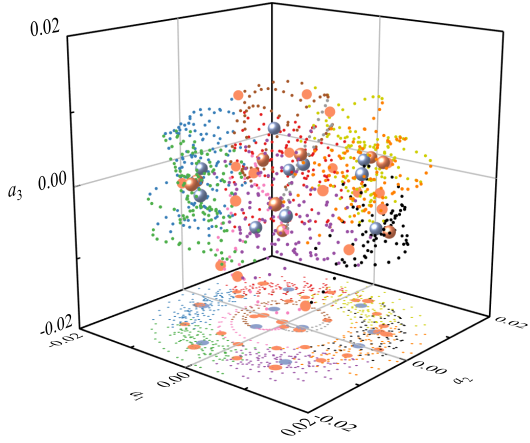}
\caption{
Same as figure~\ref{Re300_cluster3d}, but for chaotic flow at $Re = 450$.}
\label{Re450_cluster3d}
\end{figure}
\begin{figure}
\includegraphics[width=7.5cm]{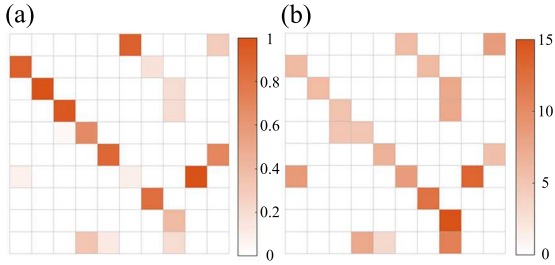}
\caption{
Same as figure~\ref{Re300_transition_probability}, but for chaotic flow at $Re = 450$. (a) Direct transition matrix $\mathbf{Q}$. (b) Transition time matrix $\mathbf{T}$.}
\label{Re450_transition_probability}
\end{figure}
\begin{figure}
\includegraphics[width=7.5cm]{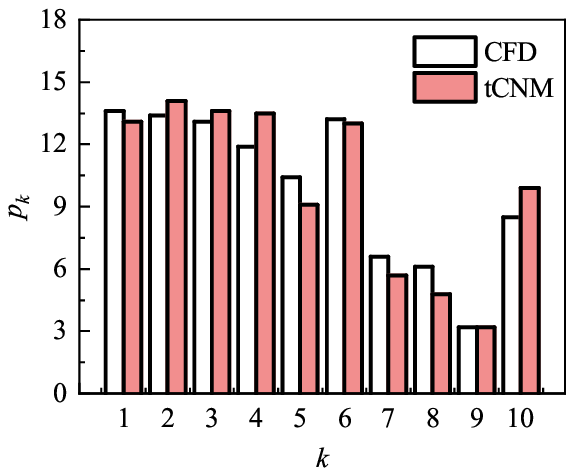}
\caption{
Same as figure~\ref{Re300_cluster_probability}, but for chaotic flow at $Re = 450$.}
\label{Re450_cluster_probability} 
\end{figure}
\begin{figure}
\includegraphics[width=7.5cm]{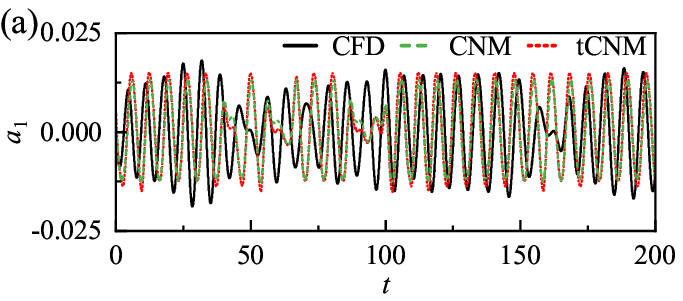}
\includegraphics[width=7.5cm]{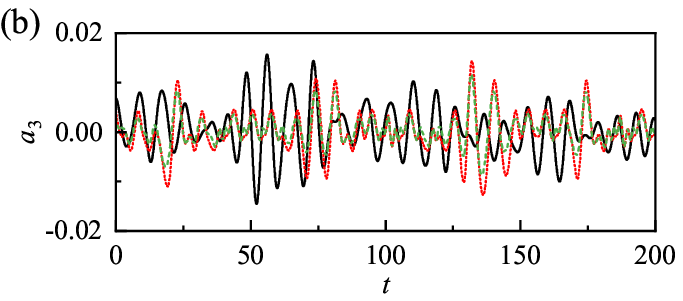}
\includegraphics[width=7.5cm]{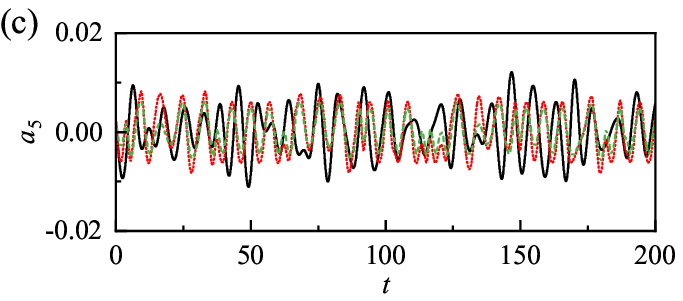}
\caption{
Same as figure~\ref{Re300_coefficients_restore}, but for chaotic flow at $Re = 450$. (a) Coefficient $a_1$. (b) Coefficient $a_3$. (c) Coefficient $a_5$.}
\label{Re450_coefficients_restore}
\end{figure}
\begin{figure}
\includegraphics[width=7.5cm]{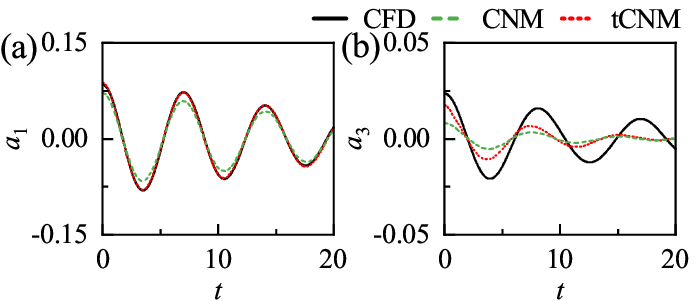}
\includegraphics[width=7.5cm]{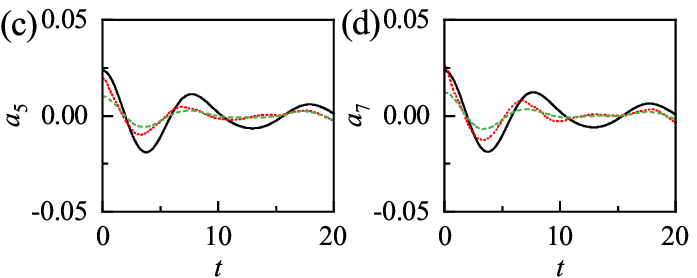}
\caption{
Same as figure~\ref{Re350_auto_correlation}, but for chaotic flow at $Re = 450$. (a) Coefficient $a_1$. (b) Coefficient $a_3$. (c) Coefficient $a_5$. (d) Coefficient $a_7$.}
\label{Re450_auto_correlation}
\end{figure}
\begin{figure*}
\includegraphics[width=\linewidth]{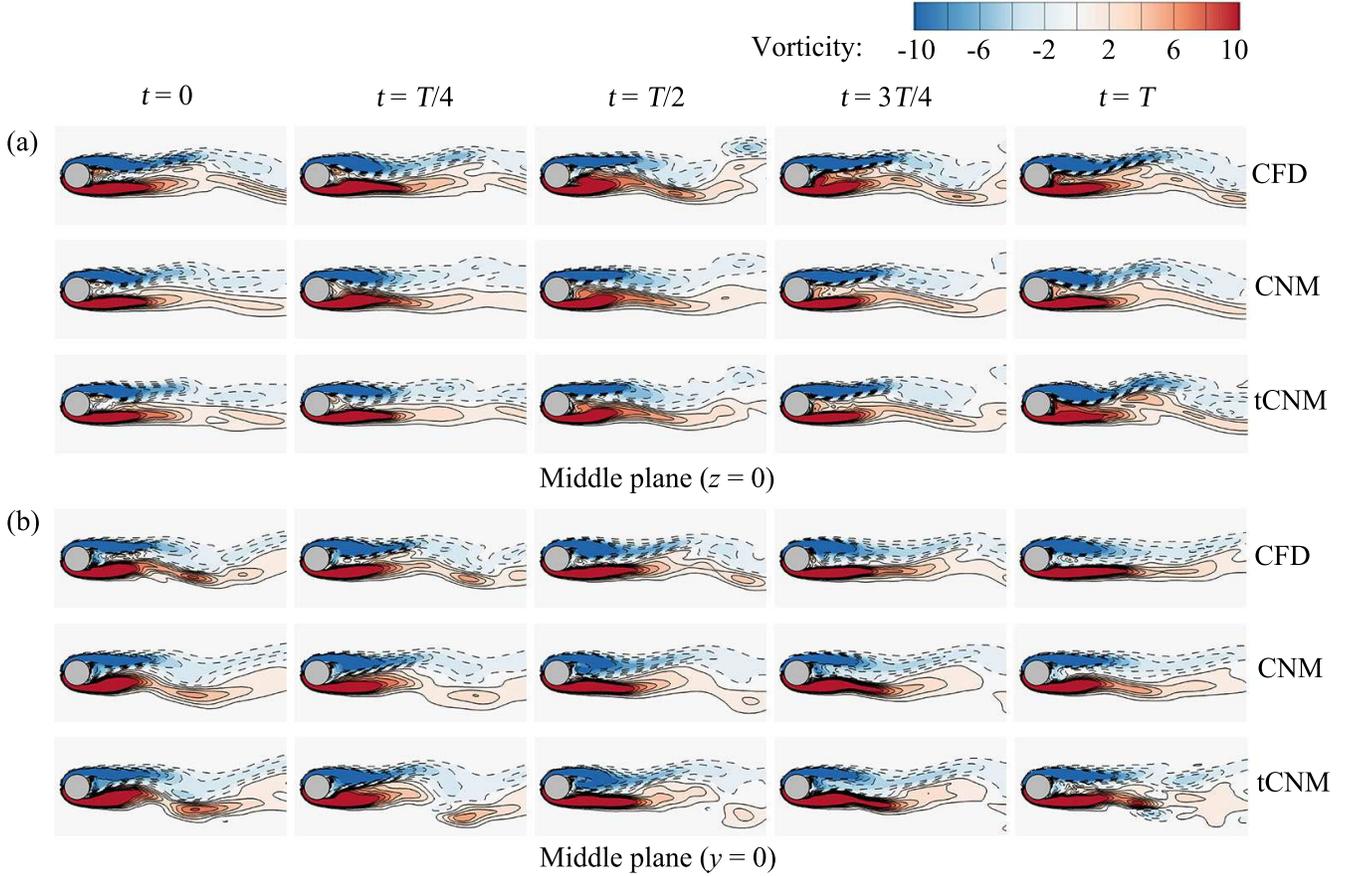}
\caption{
Same as figure~\ref{Re300_contours}, but for chaotic flow at $Re = 450$. (a) $xy$ plane at $z=0$. (b) $xz$ plane at $y=0$.}
\label{Re450_contours}
\end{figure*}

A three-dimensional view of all the snapshots and propagation nodes is shown in figure~\ref{Re450_cluster3d}.
The distribution of snapshots in the state space is rather scattered and irregular, but a ring can be seen in the top view.
The clustering algorithm divides them into different clusters, like cutting a chaotic cloud area into different colored pieces. CNM and tCNM try to figure out a clear pattern of trajectory motion statistically.
From the top view, the averaged centroids $\bm{c}_{k}^{0}$ locates deep in each piece, while the shifted cluster centroids $\bm{c}_k$ and the trajectory support points $\bm{r}_{kj}$ are near the physical trajectories.
Most of the original centroids $\bm{c}_{k}^{0}$ are away from the shifted centroids $\bm{c}_k$, and the trajectory support points will further help to refine the cluster transitions, which means that tCNM can significantly reduce the representation error of CNM for the chaotic flow.

The transition matrix $\mathbf{Q}$ and the transition time matrix $\mathbf{T}$ of chaotic flow are illustrated in figure~\ref{Re450_transition_probability}.
The transition matrix do not exist a main loop traveling through all the clusters.
The cluster transitions with high probabilities form a cycling transition from cluster $\mathcal{C}_k = 1$ to $\mathcal{C}_k = 6$, which corresponds to a vortex shedding period resolved by six clusters, appearing most frequently in this flow regime.
This also reveals the advantage of the cluster-based reduced-order modeling method in resolving the dynamics of complex dynamic systems, as a clear limit cycle can be recognized even for chaotic flow.
The transitions from the rest four clusters show to be highly random.
The cycling block from cluster $\mathcal{C}_k = 1$ to $\mathcal{C}_k = 6$ can jump into the random block from cluster $\mathcal{C}_k = 7$ to $\mathcal{C}_k = 10$ through $\mathcal{C}_k = 1, 4, 5, 6$, 
Conversely, the random clusters can enter the circular orbit from $\mathcal{C}_k = 7, 8, 10$.
$\mathcal{C}_k = 6, 8$ are the key transient points for choosing the periodic and stochastic routes, corresponding to the different shedding features.
From the time matrix $\mathbf{T}$, we can also find the dissimilarity of transition times in different transitions, as in the quasi-periodic flow. 
However, no smooth variation can be found for the random clusters.
For the clusters affiliated with the limit cycle, the transition times are similar.
For the random clusters, some transition times are much longer, indicating dense distribution of data in these clusters.

The cluster probability distribution is shown in figure~\ref{Re450_cluster_probability}.
The clusters affiliated with the limit cycle have a relatively higher distribution due to their higher transition possibility, and a uniform distribution due to their analogous transition times. Even the transition times are longer for the other clusters, they still have a lower distribution due to the lower transition probability. Note that the lower distribution for cluster $\mathcal{C}_k = 7$ attributed to its low transition probability as destination clusters from cluster $\mathcal{C}_k = 6$.
Despite the dataset being probabilistically well reconstructed, certain differences can be found between the distribution of two models and the CFD data, and even more significant than the quasi-periodic flow.
As the distribution tendency is still similar and these differences only affect the transition error, we deem these differences can have little impact in this study to compare two models with representation error.

The temporal evolution of $a_1$, $a_3$ and $a_5$ is illustrated in figure~\ref{Re450_coefficients_restore},
the autocorrelation function of $a_1$, $a_3$, $a_5$ and $a_7$ is shown by figure~\ref{Re450_auto_correlation}.
From the temporal evolution, it is clear that the amplitude of the restored coefficients by tCNM is much larger than that of CNM due to the lesser information loss.
Most of the amplitudes resolved by tCNM show a higher accuracy, similarly with the autocorrelation function.
Additionally, since the shifted centroids $\bm{c}_k$ are the averages of all the representative flow states, the trajectory of tCNM through one cluster can be seen as the average of all the optimized trajectories, which corresponds to all the slightly dissimilar vortex structures in this cluster. Thus, the extra representation error can be substantially moderated by tCNM, as demonstrated in the autocorrelation function of $a_1$ in figure~\ref{Re450_auto_correlation}.
For the higher-order mode coefficients, the fluctuation level of the autocorrelation function for both tCNM and CNM is lower than the CFD data, which is mainly caused by the transition error.

The vortex distribution on the middle $xy$ and $xz$ plane are illustrated in figure~\ref{Re450_contours}.
For chaotic flow, as the separation point is rotating around the sphere and the symmetric plane is lost, we did not do the rotation to the numerical domain.
The contours of CNM are like the averaged flow state nearly in every snapshot, and the coherent structures and secondary vortexes cannot be easily recognized compared with tCNM.
Meanwhile, slight deformation also occurs on the vortex pair. 
For tCNM, the largest differences compared with the CFD data still occurs on the first and last snapshots, the reason is the same as the quasi-periodic flow.
While for the other snapshots, most of the delicate vortex structures are well preserved, which is explained as the consequence of the combined improvement by centroid shift and inter-cluster constraint.
\begin{figure}
\includegraphics[width=7.5cm]{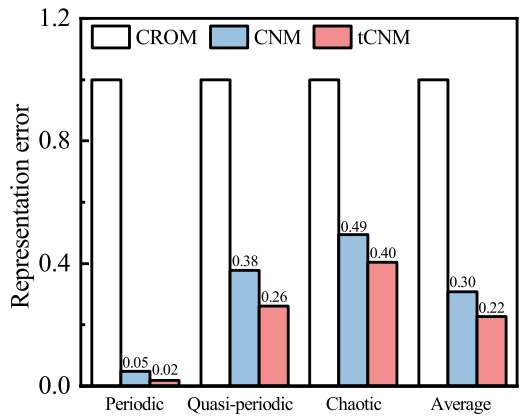}
\caption{Representation error of three cluster-based models for the three flow regimes and the corresponding average. All the levels are normalized with respect to the representation error of CROM.}
\label{Representation error}
\end{figure}

In figure~\ref{Representation error}, to illustrate the performance improvement of tCNM, a statistical comparison based on the representation error between three cluster-based models is conducted for all the three flow regimes.
The representation error evaluates the reconstructed dynamics to the raw snapshots.
For tCNM and CNM, this error is defined as the average of the distances from the snapshot to the reconstructed trajectory. Whereas for CROM, the dynamics is expressed as direct jumps between discrete cluster centroids. Hence this error is defined as the average of the distances from the snapshots to their nearest centroids.
The results of the three models are normalized by the error of CROM for intuitive comparison.
CNM and tCNM achieve a significant improvement in the representation accuracy by counting the state propagation of the network model, especially for the periodic flow. 
The improvement for the other flow regimes is relatively unapparent, mainly due to the increasing intra-cluster variance for the more complex flow features.
Compared with CNM, the representation errors of tCNM on periodic flow, quasi-periodic flow, and chaotic flow are reduced by $60\%$, $31\%$, and $18\%$, respectively, with an average of $27\%$.

\section{Conclusions}
\label{sec6}

In this work, we propose a novel data-driven reduced-order modeling methodology for time-resolved data.
The starting point is the Cluster-based Network Model (CNM) \citep{2021_SCIA_fernex_cluster, 2021_JFM_li_cluster}. 
The $k$-means++ algorithm optimizes the representation error of the snapshots by the closest centroids. \citet{2014_JFM_kaiser_cluster} proposed a Cluster-based Markov Model to analyze the flow dynamics between these representative centroids.
Yet, the CNM includes time-continuous `flights' between neighboring centroids and builds a directed graph emphasizing the non-trivial transitions between nodes in the network. 
The key innovation of the proposed trajectory-optimized Cluster-based Network Model (tCNM) is to shift the centroids and allow constraints between centroids in order to decrease the kinematic representation error of interpolated states and thus to increase the accuracy of the reconstructed dynamics.
In particular, 
the snapshots are coarse-grained into a few clusters following the $k$-means++ algorithm.
Then, the centroids $\bm{c}_k$ are shifted closer to the snapshot trajectory, meanwhile, the trajectory support points $\bm{r}_{kj}$ are built in order to refine the transitions between two cluster centroids.

The resulting tCNM has several desirable features: 
\begin{enumerate}
    \item tCNM retains all the advantages of CNM. 
    The framework does not require any first-principle knowledge.
    It can be extended to incorporate multiple operating conditions or control-oriented predictions beyond the training data .
    \item The representation error is reduced, without losing simplicity and robustness. The propagation trajectory is more accurate, especially near the critical moment when the vortex is about to generate or separate.
    \item The off-line computational load is only slightly larger than CNM. The extra computations are mainly due to the building of the new nodes, which requires only a tiny fraction of total computational operations.
    \item The methodology is data-driven and fully automatable, including the centroid shift and the inter-cluster constraint.
\end{enumerate}

However, tCNM, or any other data-driven model order reduction,
requires a careful calibration of the observation domain.
If the domain is too small, future states are difficult to predict
and challenge the transition model.
If, on the other hand, the domain comprises too many uncorrelated flow structures, the number of necessary centroids  becomes excessive.
Also, the required snapshot data will become impractically large,
as each centroid represents a coarse-grained recurrent state 
which should be found in multiple snapshots.
We note that similar observations also hold for POD.

The three-dimensional wake behind a sphere of three different dynamics is considered for the evaluation of tCNM:
(1) The periodic flow features a periodic shedding of hairpin vortices at $Re=300$.
(2) The quasi-periodic flow exhibits an irregular shedding with a long-time periodicity at $Re=350$. 
(3) The chaotic flow shows a fully three-dimensional irregular shedding with the separation point slowly rotating around the sphere at $Re=450$.
For all three cases, the snapshots are coarse-grained into $K=10$ clusters yielding satisfactory dynamics predictions.

On a  kinematic level, 
clustering discretizes the flow states.
Each centroid $\bm{c_k^0}$ represents realizable states modulo a coarse graining.
In fact,  the  $K$ centroids constitute a complete set of probable and recurrent states with distinct vortex formations.
The proposed tCNM shifts of the centroids improve the representation error to POD level.
In contrast,  linear combinations of POD modes are needed to approximate flows, and the individual modes typically consist of a mixture of frequencies. Both features mitigate their interpretability.
On a dynamic level, the time-relevant dynamics between the snapshots are represented by transitions between different clusters, or, in other words between different centroids.
These transitions illustrated by the transition matrixes serve as a deterministic–stochastic gray-box model resolving the coherent-structure evolution,
e.g., the deterministic limit cycle for a periodic flow regime; the stochastic model for a chaotic flow regime, which chooses the next destination according to the probability of transitions.    
Hence, CNM and tCNM can automatically capture the dynamics of the unsteady flow field, while POD models can only capture the energy-dominate modes and require additional nonlinear modeling for the dynamics.
Even though CNM well resolves the temporal evolution of the flow dynamics, a noteworthy deficiency can be seen in dealing with the representation error. 
However, compared with CNM, tCNM achieved a reduction of $27\%$ for the representation error on average of the three flow regimes.

tCNM opens a novel automatable avenue for data-driven nonlinear reduced-order modeling.
tCNM enables cluster-based methods with optimized trajectory to enhance linear interpolation, thereby reducing representation error and improving the accuracy of the dynamical model.
The methodology can be applied to numerous other applications and extended to parametric and control-oriented reduced-order models. 

As outlook, 
tCNM foreshadows the advantages of a new optimization paradigm of clustering: 
the centroids may be chosen to minimize representation errors with respect to local interpolations.
In addition, the trajectory models may be further refined with respect to historical data, 
like for the higher-order Markov chain model \citep{2021_SCIA_fernex_cluster}.
The authors are actively pursuing these avenues.

\begin{acknowledgments}

Chang Hou appreciates the support from the School of Mechanical Engineering and Automation during his PhD thesis at Harbin Institute of Technology (Shenzhen).
This work is supported by the National Natural Science Foundation of China (NSFC) under grant 12172109 and 12172111, 
and by the Natural Science and Engineering grant 2022A1515011492 of Guangdong province, China.
We appreciate valuable discussions with Guy Y. Cornejo-Maceda, 
Hao Li, Richard Semaan and the HIT-Hanghua team. 
Last but not least, we thank the referees 
for their many constructive suggestions.

The authors declare that there is no conflict of interest regarding the publication of this paper.
\end{acknowledgments}

\section*{Appendix: Cluster centroids versus POD modes}
\begin{figure}[!h] 
\includegraphics[width=7.5cm]{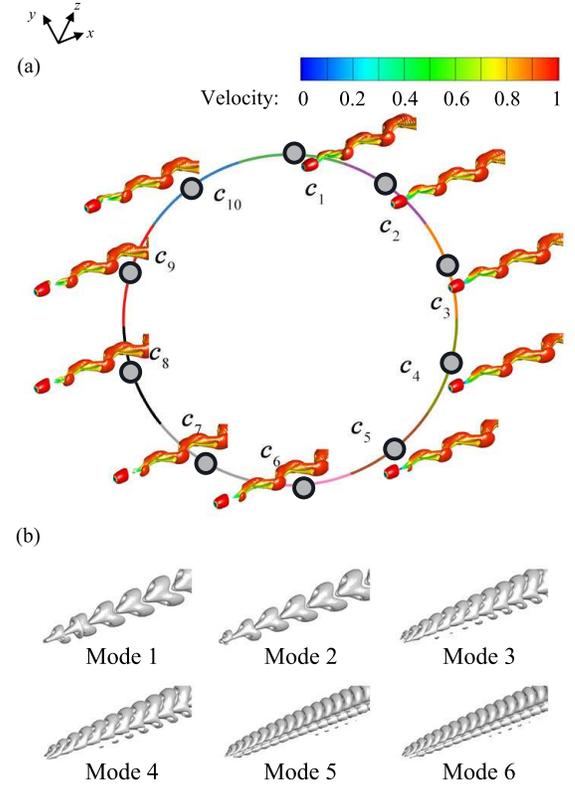}
\caption{Comparison between the tCNM centroids and the first POD modes for the periodic flow at $Re = 300$. 
(a) Graph of transitions between clusters. 
The centroids are marked with large gray dots in the proximity map, and their vortex structures are visualized like in figure figure~\ref{Vortex_structure}. 
The snapshots belonging to different clusters are marked with small dots in different colors. 
(b) The leading 6 POD modes. 
The vortex structures are visualized by the $Q$-criterion ($Q = 0.001$).}
\label{C_M_300}
\end{figure}
\begin{figure}[!h] 
\includegraphics[width=7.5cm]{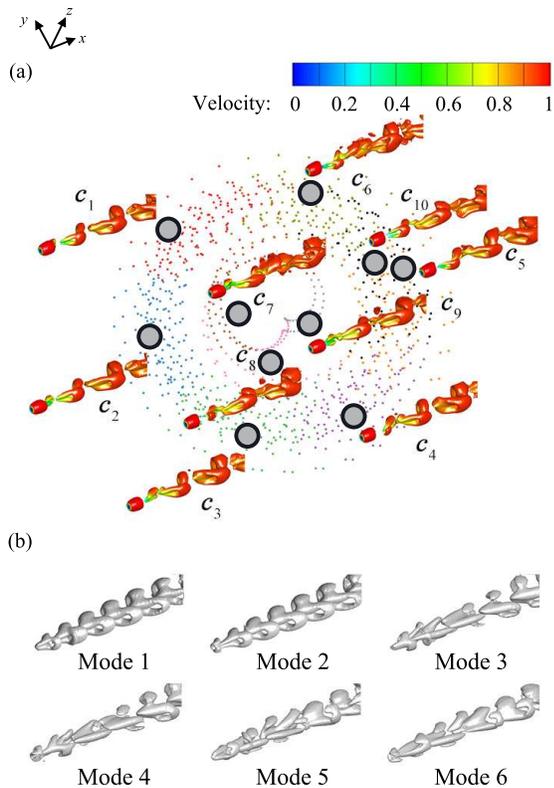}
\caption{Same as figure~\ref{C_M_300}, but for chaotic flow at $Re = 450$. (a) Graph of transitions between clusters. (b) The leading 6 POD modes.}
\label{C_M_450}
\end{figure}
The tCNM centroids and the leading 6 POD modes are compared in figure~\ref{C_M_300} for periodic flow and in figure~\ref{C_M_450} for the chaotic flow of the sphere wake.
After clustering, 
the time-resolved snapshots of the sphere wake are grouped into different clusters depending on their flow characteristics.
Each cluster is represented by its centroid.
For tCNM, these snapshots are shifted towards the trajectory and can therefore be seen as more realistic flow states.
In contrast, POD modes come with a mathematically optimal representation error but are typically not physically interpretable flow states:
First, POD needs the superposition of individual modes to approximate the snapshot data.
Second, POD modes typically comprise multiple frequencies, making the physical interpretation difficult.
Third, POD modes represent a first and second-moment statistical analysis ignoring the temporal dynamics.
The last two features might be improved with other empirical Galerkin expansions 
like DMD \citep{Rowley2009jfm,Schmid2010jfm}, 
recursive DMD \citep{2016_JFM_noack_recursive}, 
spectral \citep{Sieber2016jfm}, etc. by sacrificing the optimal representation error.

\section*{Data Availability Statement}
The data that support the findings of this study are available from the first author upon reasonable request.

\bibliography{reference,all}

\end{document}